\documentclass[prd,twocolumn,floatfix,amsmath,nofootinbib,amssymb,floatfix]{revtex4}
\usepackage{graphicx,color,dcolumn,booktabs,bm}
\usepackage{longtable,lscape}
\usepackage{pdfpages}
\usepackage{txfonts}
\usepackage{overpic}
\usepackage{amssymb}
\usepackage{makecell}
\usepackage{indentfirst}
\usepackage{feynmf}   %{feynmp}
\usepackage{slashed}  %for Feynman symbols
\usepackage{cases}
\usepackage{color}
\usepackage{multirow}
\usepackage{threeparttable}
\usepackage{epstopdf}
\usepackage{enumerate}
\usepackage{subfigure}
\usepackage{diagbox}
\usepackage{graphicx,color,dcolumn,booktabs,bm}
\usepackage{mathrsfs}
\usepackage{cancel}
\usepackage{float}
\usepackage[misc]{ifsym}
\usepackage[colorlinks,
citecolor=blue,
anchorcolor=red,
menucolor=red,
linkcolor=red,
filecolor=red,
runcolor=red,
urlcolor=blue,
frenchlinks=red]{hyperref}
\usepackage{array}
\usepackage{booktabs}

\begin{document}
\title{Analysis of the strong vertices $\Lambda_cD^{(*)}N^*(1535)$ and $\Lambda_bB^{(*)}N^*(1535)$ in QCD sum rules}
\author{Jie Lu$^{1,2}$}
\email{l17693567997@163.com}
\author{Dian-Yong Chen$^{1,3}$}
\email{chendy@seu.edu.cn}
\author{Guo-Liang Yu$^{2}$}
\email{yuguoliang2011@163.com}
\author{Zhi-Gang Wang$^{2}$}
\email{zgwang@aliyun.com}
\author{Ze Zhou$^{2}$}

\affiliation{$^1$ School of Physics, Southeast University, Nanjing 210094, People's Republic of
China\\$^2$ Department of Mathematics and Physics, North China
Electric Power University, Baoding 071003, People's Republic of
China\\$^3$ Lanzhou Center for Theoretical Physics, Lanzhou University, Lanzhou 730000, People's Republic of
China}
\date{\today }

\begin{abstract}

In this article, we firstly analyze the mass and pole residue of negative parity nucleon $N^*(1535)$ within the two-point QCD sum rules. Basing on these results, we continuously study the strong coupling constants of vertices $\Lambda_cDN^*$, $\Lambda_cD^*N^*$, $\Lambda_bBN^*$ and $\Lambda_bB^*N^*$ in the framework of three-point QCD sum rules. At hadron side, all possible couplings of interpolating current to hadronic states are considered. At QCD side, the contributions of vacuum condensate terms $\langle\bar{q}q\rangle$, $\langle g_s^2GG\rangle$, $\langle\bar{q} g_s\sigma Gq\rangle$, $\langle\bar{q}q\rangle^2$ and $g_s^2\langle\bar{q}q\rangle^2$ are also considered. By setting the four momentum of $D^{(*)}[B^{(*)}]$ mesons off-shell, the strong coupling constants in deep space-like regions ($Q^2=-q^2\gg\Lambda_{QCD}^2$) are obtained. Then, the momentum dependent coupling constants in space-like regions are fitted into analytical function $G(Q^2)$ and are extrapolated into time-like regions ($Q^2<0$). Finally, the on-shell values of strong coupling constants are obtained by taking $Q^{2}=-m_{D^{(*)}[B^{(*)}]}^2$. The results are $G_{\Lambda_cDN^*}(Q^2=-m_D^2)=4.06^{+0.96}_{-0.75}$, $f_{\Lambda_cD^*N^*}(Q^2=-m_{D^*}^2)=3.73^{+0.68}_{-0.16}$, $g_{\Lambda_cD^*N^*}(Q^2=-m_{D^*}^2)=9.22^{+3.16}_{-0.36}$,
$G_{\Lambda_bBN^*}(Q^2=-m_B^2)=9.11^{+1.54}_{-1.61}$,
$f_{\Lambda_bB^*N^*}(Q^2=-m_{B^*}^2)=8.55^{+2.69}_{-2.21}$ and
$g_{\Lambda_bB^*N^*}(Q^2=-m_{B^*}^2)=-0.25^{+0.16}_{-0.01}$. 

\end{abstract}

\pacs{13.25.Ft; 14.40.Lb}

\maketitle

\section{Introduction}\label{sec1}

The study of interactions between hadrons has always been an important issue in quantum chromodynamics (QCD) which is the fundamental theory describing the strong interaction. Although QCD has achieved remarkable success in the high energy regions, it has non-perturbative properties at low energy regions which makes it difficult to understand the interactions in hadronic level, especially in systems involving heavy quarks. As key parameters, the strong coupling constants are important in both theoretical researches and experimental observations which can help us to study the physical quantities such as decay widths and production cross sections.

The charmed (bottom) hadrons which contain a heavy quark $c$ or $b$ is an excellent laboratory to study the interaction between heavy and light quarks. For example, the charmed (bottom) mesons such as $D^{(*)}$, $B^{(*)}$ and charmed (bottom) baryons such as $\Lambda_{c[b]}$, $\Sigma_{c[b]}$, exhibit distinct behaviors due to the heavy quark masses. As for the nucleon resonances with negative parity such as $N^*(1535)$, $N^*(1520)$, they play important roles in nuclear physics which serving as gateways to understanding the excitation state spectrum of baryons and their decay properties. The research of the coupling constants between these nucleon resonance and charmed (bottom) hadrons is also interesting. These parameters can provide useful information to determined the spin and inner structure of hadrons and to understand the breaking of chiral and heavy quark symmetry and to predict the existence of exotic states such as pentaquark.

It is very difficult to study the strong coupling constants from the QCD first principle. However, as important parameter, the strong coupling constants are urgently needed in studying the production and decay processes of traditional and exotic hadron states~\cite{Xiao:2019mvs,He:2019rva,Wu:2024lud}. In addition, the  long-distance dynamics in the decay processes of heavy flavor hadron can be studied using the final state re-scattering mechanism~\cite{Han:2021azw,Jia:2024pyb,Hu:2024uia}. In this picture, the interaction can be described as some hadron loops where the strong coupling constants are important parameters. A general way to determine the strong coupling constant is to use the effective field theory to obtain the theoretical expression about the decay width or production cross section, and then to determine the value of the coupling constant according to the experimental results of the decay width or cross section. However experimental data is limited, and some decay and production processes related to strong vertices can not occur directly. Therefore, many non-perturbative approaches are employed to study the strong coupling constants, such as lattice QCD~\cite{Altmeyer:1995qx}, QCD sum rules~\cite{Bracco:1999xe,Bracco:2010bf,Bracco:2011pg,Cui:2012wk,Wang:2013iia,Yu:2015xwa,Lu:2023gmd,Lu:2023lvu}, Light-cone QCD sum rules~\cite{Colangelo:1997rp,Zhu:1998vf,Khodjamirian:1999hb,Wang:2006bs,Wang:2007mc,Khodjamirian:2011jp,Khodjamirian:2020mlb,Aliev:2020lly,Aliev:2021hqq,Aliev:2022gxi} and other methods~\cite{Oh:2000qr,Li:2002pp,Deandrea:2003pv}.

As a powerful non-perturbative method, the QCD sum rules has been widely used in studying the properties of hadrons~\cite{Shifman:1978bx,Shifman:1978by}. Specifically, the QCD sum rules based on two-point correlation function are used to analyze the mass, decay constant and distribution amplitude of hadrons~\cite{Novikov:1977dq,Ioffe:1981kw,Narison:1987qc,Matheus:2009vq,Mo:2014nua,Wang:2015mxa,Wang:2020mxk,Yang:2023fsc,Zeng:2025gft} and three-point QCD sum rules are used to analyze the hadron transition form factors and strong coupling constants between hadrons~\cite{Wang:2007ys,Azizi:2015tya,Azizi:2015bxa,Yu:2016pyo,Yu:2018hnv,Shi:2019hbf,Zhao:2020mod,Zhao:2021sje,Zhang:2023nxl,Lu:2023pcg,Wu:2024gcq,Zhang:2024ick,Lu:2025bvi}. In our previous works, the strong coupling constants of vertices $\Sigma_cD^{(*)}N$, $\Sigma_bB^{(*)}N$, $\Sigma_cD^*\Delta$ and $\Sigma_bB^*\Delta$ have been analyzed within three-point QCD sum rules~\cite{Yu:2016pyo,Yu:2018hnv,Lu:2023pcg}. As a continuation of these works, we analyze the vertices $\Lambda_cD^{(*)}N^*$ and $\Lambda_bB^{(*)}N^*$ in this article where all the contributions of positive (negative) parity baryons are considered.

This article is organized as follows. After introduction in Sec. \ref{sec1}, we firstly analyze the mass and pole residue of negative parity nucleon $N^*(1535)$ with the two-point QCD sum rules in Sec. \ref{sec2}. Then, we analyze the strong coupling constants of vertices $\Lambda_cDN^*$, $\Lambda_cD^*N^*$, $\Lambda_bBN^*$ and $\Lambda_bB^*N^*$ with the three-point QCD sum rules in Sec. \ref{sec3}. Sec. \ref{sec4} is employed to present the numerical results and discussions. Sec. \ref{sec5} is the conclusion part. Some complicated formulas are shown in Appendix~\ref{Sec:AppA}.

\section{Two-point QCD sum rules for mass and pole residue}\label{sec2}

There have been lots of literatures to study the decay constants of charmed (bottom) mesons and pole residues of charmed (bottom) baryons~\cite{Narison:1987qc,Wang:2015mxa,Wang:2020mxk,Zhao:2020mod}. However, the study about pole residues of nucleon resonance with negative parity is limited. Thus, we firstly analyze the mass and pole residue of negative parity nucleon $N^*(1535)$ by two-point QCD sum rules in this section.

\subsection{The interpolating currents and two-point correlation function}\label{sec2.1}
To construct the QCD sum rules, one should choose appropriate interpolating quark currents for corresponding hadrons. For charmed (bottom) baryons $\Lambda_{c[b]}$, we adopt the following interpolating current in present work,
\begin{eqnarray}\label{eq:1}
J_{\Lambda _{c[b]}}(x) &&= \varepsilon _{ijk}\left( {u^{iT}(x)\mathcal{C}\gamma _5d^j(x)} \right)c[b]^k(x)
\end{eqnarray}
where $\varepsilon_{ijk}$ is the 3 dimension Levi-Civita tensor, $i$, $j$ and $k$ represent the color indices, and $\mathcal{C}$ is the charge conjugation operator.

The $N^{*}(1535)$ have $J^{P}$ quantum numbers $\frac{1}{2}^{-}$, and can be regarded as the negative partner of nucleons $p$ or $n$ with positive parity. Therefore, their interpolating currents can be simply expressed as the Ioffe current of $p$ multiplied by dirac matrix $\gamma_{5}$~\cite{Ioffe:1981kw}.
\begin{eqnarray}\label{eq:2}
J(x) &&= \varepsilon _{ijk}\left( u^{iT}(x)\mathcal{C}\gamma _\beta u^j(x) \right)\gamma _\beta d^k(x)
\end{eqnarray}

The two-point correlation function can be defined as,
\begin{eqnarray}\label{eq:3}
\Pi(p') &&= i\int d^4xe^{ip'x}\left\langle 0 \right| \mathcal{T}\{ J(x)\bar J(0)\} \left| 0 \right\rangle
\end{eqnarray}
where $p'$ is the four momentum of the virtual negative parity nucleon with $p'^2<0$, $\mathcal{T}$ denotes the time ordered product and $\bar{J}(0)=J^{\dag}(0)\gamma_{0}$.

In the framework of QCD sum rules, the above correlation function can be handled at hadron and quark levels respectively, the former is called the phenomenological side and the latter is called the QCD side. Finally, we can match the calculation of these two levels by quark-hadron duality condition and obtain the QCD sum rules for the parameters of hadrons.

\subsection{The phenomenological side}\label{sec2.2}

In the phenomenological side, we insert complete sets of hadron states with the same quantum numbers as the interpolating current into the correlation function. After finishing the integrals in coordinate space, isolating the contributions of the ground states from excited states and using the dispersion relation, the expression of this two-point correlation function can be written as,
\begin{eqnarray}\label{eq:4}
\notag
\Pi^{\mathrm{phy}}(p') &&= \frac{\left\langle 0 \right|J(0)\left| N^*(p') \right\rangle \left\langle N^*(p') \right|\bar J(0)\left| 0 \right\rangle }{ m_{N^*}^2-p'^2}\\
&& + \frac{\left\langle 0 \right|J(0)\left| p(p') \right\rangle \left\langle p(p') \right|\bar J(0)\left| 0 \right\rangle }{m_p^2-p'^2} + ...
\end{eqnarray}
where ellipsis represents the contributions of higher resonances and continuum states. From this above equation, one can find that the current $J(0)$ not only couples to $N^{*}(1535)$ ($J^{P}=\frac{1}{2}^{-}$), but also to the proton $p$ ($J^{P}=\frac{1}{2}^{+}$). In Eq. (\ref{eq:4}), the vacuum matrix elements are defined as follows,
\begin{eqnarray}\label{eq:5}
\notag
&&\left\langle 0 \right|J(0)\left| N^*(q) \right\rangle  = \lambda _{N^*}u(p',s)\\
&&\left\langle 0 \right|J(0)\left| p(p') \right\rangle  = -\lambda _p\gamma _5u(p',s)
\end{eqnarray}
where $\lambda_{N^*}$ and $\lambda_{p}$ are the pole residues of the baryons $N^*(1535)$ and $p$, respectively. The $u(p',s)$ denotes the spinor wave functions of $J^P=\frac{1}{2}^\pm$ nucleons which satisfy the following spin polarization summation formula,
\begin{eqnarray}\label{eq:6}
\sum\limits_{s} u(p',s)\bar u(p',s) = \slashed p' + m_{N^*[p]}
\end{eqnarray}
With Eqs. (\ref{eq:5}) and (\ref{eq:6}), the correlation function in Eq. (\ref{eq:4}) can be written as,
\begin{eqnarray}\label{eq:7}
\notag
\Pi ^{\mathrm{phy}}(p') &&= \Pi _1^{\mathrm{phy}}(p'^2)\slashed p' + \Pi _2^{\mathrm{phy}}(p'^2)\\
\notag
&& = \left(\frac{\lambda _{N^*}^2}{m_{N^*}^2-p'^2} + \frac{\lambda _p^2}{ m_p^2-p'^2}\right)\slashed p'\\
&& + \left(\frac{\lambda _{N^*}^2m_{N^*}}{m_{N^*}^2-p'^2} - \frac{\lambda _p^2m_p}{m_p^2-p'^2}\right) + ...
\end{eqnarray}

\subsection{The QCD side}\label{sec2.3}

In the QCD side, we substitute the interpolating current in Eq. (\ref{eq:3}) with Eq. (\ref{eq:2}), and do the operator product expansion (OPE) by contracting the quark filed with Wick's theorem. After these processes, the correlation function in QCD side can be expressed as,
\begin{eqnarray}\label{eq:8}
\notag
\Pi ^{\mathrm{QCD}}(p') &&= 2i\varepsilon _{ijk}\varepsilon _{i'j'k'}\int d^4x e^{ip'x}\\
&& \times Tr[U^{jk'}(x)\gamma _\alpha \mathcal{C}U^{j'iT}(x)\mathcal{C}\gamma _\beta]\gamma _\beta D^{ki'}(x)\gamma _\alpha
\end{eqnarray}
where, $U^{ki'}(x)$ and $D^{k'j}(x)$ are $u$ and $d$ quark full propagators and they can be uniformly expressed as the following form in coordinate space~\cite{Pascual:1984zb,Reinders:1984sr},
\begin{eqnarray}\label{eq:9}
\notag
U^{ij}(x) &&= D^{ij}(x) = \frac{i\delta ^{ij}\slashed{x}}{2\pi ^2x^4} - \frac{\delta ^{ij}\left\langle \bar qq \right\rangle }{12} - \frac{\delta ^{ij}x^2\left\langle \bar qg_s\sigma Gq \right\rangle }{192}\\
\notag
&& - \frac{\left\langle {\bar q^j\sigma ^{\mu \nu }q^i} \right\rangle \sigma _{\mu \nu }}{8} - \frac{ig_sG_{\alpha \beta }^nt_{ij}^n(\slashed{x}\sigma ^{\alpha \beta } + \sigma ^{\alpha \beta}\slashed x)}{32\pi ^2x^2}\\
&& - \frac{i\delta ^{ij}x^2\slashed{x}g_s^2\left\langle \bar qq \right\rangle ^2}{7776} + ...
\end{eqnarray}
Here, $\left\langle \bar qg_s\sigma Gq \right\rangle=\left\langle \bar qg_s\sigma_{\mu\nu}t^\alpha G^\alpha_{\mu\nu}q \right\rangle$, $t^{\alpha}=\frac{\lambda^\alpha}{2}$, $\lambda^{\alpha}$ ($\alpha=1,...,8$) are the Gell-Mann matrices, and $\sigma_{\mu\nu}=\frac{i}{2}[\gamma_\mu,\gamma_\nu]$.

According to the dispersion relation, the correlation functions in QCD side can be expressed as the following form,
\begin{eqnarray}\label{eq:10}
\Pi_i ^{\mathrm{QCD}}(p') &&= \int\limits_{u_{min}}^\infty  du\frac{\rho _i^{\mathrm{QCD}}(u)}{u - p'^2}
\end{eqnarray}
where, $\rho^{\mathrm{QCD}}_i(u)$ is QCD spectral density and it can be obtained by taking the imaginary part of the correlation function,
\begin{eqnarray}\label{eq:11}
\rho _i^{\mathrm{QCD}}(u) &&= \frac{1}{\pi}{\mathop{\rm Im}\nolimits} [\Pi _i^{\mathrm{QCD}}(u+i\epsilon)]
\end{eqnarray}
Here $u=p'^2$ in Eq.~(\ref{eq:10}), $u_{\min}$ is commonly called the creative threshold and its value usually taken as the square of the summation of the quark masses that make up the hadron state. For the hadron composed of $u$ and $d$ quarks, the $u_{\min}$ can be take as zero because of the small masses of $u$ and $d$ quarks. The full expressions of these QCD spectral density are as follows,
\begin{eqnarray}\label{eq:12}
\notag
\rho _1^{\mathrm{QCD}}(u) &&= \frac{u^2}{64\pi ^4} + \frac{\left\langle g_s^2GG \right\rangle }{128\pi ^4} + \left[ \frac{2\left\langle \bar qq \right\rangle ^2}{3} + \frac{4\alpha _s\left\langle {\bar qq} \right\rangle ^2}{27\pi } \right.\\
\notag
&&\left.  - \frac{\left\langle \bar qq \right\rangle \left\langle \bar qg_s\sigma Gq \right\rangle }{3}\frac{1}{T^2} \right]\delta (u)\\
\rho _2^{\mathrm{QCD}}(u) &&= \frac{\left\langle \bar qq \right\rangle }{4\pi ^2}u - \frac{\left\langle \bar qg_s\sigma Gq \right\rangle}{8\pi ^2} - \frac{\left\langle \bar qq \right\rangle \left\langle g_s^2GG \right\rangle}{144\pi ^2}\delta (u)
\end{eqnarray}

\subsection{The mass and pole residue for $N^*$(1535)}\label{sec2.4}

Taking the change of variables $p'^2\to -P'^2$, performing the Borel transformation~\cite{Reinders:1984sr} to both phenomenological and QCD sides and matching these two sides by quark-hadron duality condition, the QCD sum rules for $N^{*}(1535)$ are obtained. By using some mathematical tricks, the mass and pole residue of $N^{*}(1535)$ can be expressed as follows,
\begin{eqnarray}\label{eq:13}
\notag
m_{N^*}^2 &&= \frac{\int\limits_{u_{\min}}^{u_0} due^{- u/T^2} u\left[m_p\rho _1^{\mathrm{QCD}}(u) + \rho _2^{\mathrm{QCD}}(u)\right]}{\int\limits_{u_{\min}}^{u_0} due^{ - u/T^2} \left[m_p\rho _1^{\mathrm{QCD}}(u) + \rho _2^{\mathrm{QCD}}(u)\right]}\\
\lambda _{N^*}^2 &&= \frac{\int\limits_{u_{\min}}^{u_0} due^ {- u/T^2} \left[m_p\rho _1^{\mathrm{QCD}}(u) + \rho _2^{\mathrm{QCD}}(u)\right]e^{m_{N^*}^2/T^2}}{m_{N^*} + m_p}
\end{eqnarray}
where $T$ is the Borel parameter. $u_{0}$ is the threshold parameter which is introduced to eliminate the contributions of higher resonances and continuum states. It commonly fulfills the relation $\sqrt{u_0}=m_{\mathrm{ground}}+\delta$, where the $m_{\mathrm{ground}}$ denotes the mass of ground state hadron, $\delta$ is the energy gap between the ground and first excited states and commonly taken as a value of $0.3-0.7$ GeV which is based on experimental data and previous QCDSR calculation. For example, the mass gaps between $\eta_c(1S)$ ($J/\psi(1S)$) and $\eta_c(2S)$ ($J/\psi(2S)$) are about $0.66 (0.60)$ GeV~\cite{ParticleDataGroup:2024cfk} and the mass gap of $1{}^3{D_1}$ and $2{}^3{D_1}$ $\rho$ mesons is predicted to be about 0.3 GeV~\cite{Yu:2021ggd}. In Eq. (\ref{eq:13}), we take the mass of positive nucleon as an input parameter to eliminate the contributions of positive parity nucleon and to get the QCD sum rules for negative parity nucleons.

\section{Three-point QCD sum rules for strong coupling vertices}\label{sec3}

In this section, we give a detailed analysis for the strong coupling constants of vertices $\Lambda_cD^{(*)}N^*$ and $\Lambda_bB^{(*)}N^*$ in the framework of three-point QCD sum rules.

\subsection{The strong vertices and three-point correlation functions}\label{sec3.1}

The couplings about nucleons, pseudoscalar or vector charmed (bottom) mesons and lambda-type single charmed (bottom) baryons can be described by the following effective Lagrangian,
\begin{eqnarray}\label{eq:14}
\notag
	\mathcal{L} &&= iG_{\Lambda _{c[b]}D[B]N^*}N^*D[B]\bar \Lambda _{c[b]} + iG_{\Lambda _{c[b]}D[B]p}p\gamma _5D[B]\bar\Lambda _{c[b]}\\
	\notag
	&&+iG_{\Lambda^*_{c[b]}D[B]N^*}N^*\gamma_5D[B]\bar \Lambda^*_{c[b]} + iG_{\Lambda^*_{c[b]}D[B]p}pD[B]\bar\Lambda^*_{c[b]}\\
	\notag
	&&+N^*\left( f_{\Lambda _{c[b]}D^*[B^*]N^*}\gamma _\beta  - \frac{ig_{\Lambda _{c[b]}D^*[B^*]N^*}}{m_{\Lambda _{c[b]}} + m_{N^*}}\sigma _{\beta \alpha}q^\alpha \right)D[B]^{*\beta}\gamma _5\bar \Lambda _{c[b]}\\
	\notag
	&&+ p\left( f_{\Lambda _{c[b]}D^*[B^*]p}\gamma _\beta - \frac{ig_{\Lambda _{c[b]}D^*[B^*]p}}{m_{\Lambda _{c[b]}} + m_p}\sigma _{\beta \alpha }q^\alpha \right)D[B]^{*\beta }\bar \Lambda_{c[b]} \\
	\notag
	&&+N^*\left( f_{\Lambda^*_{c[b]}D^*[B^*]N^*}\gamma _\beta  - \frac{ig_{\Lambda^* _{c[b]}D^*[B^*]N^*}}{m_{\Lambda^*_{c[b]}} + m_{N^*}}\sigma _{\beta \alpha}q^\alpha \right)D[B]^{*\beta}\bar \Lambda^*_{c[b]}\\
	\notag
	&&+ p\left( f_{\Lambda^*_{c[b]}D^*[B^*]p}\gamma _\beta - \frac{ig_{\Lambda^*_{c[b]}D^*[B^*]p}}{m_{\Lambda^*_{c[b]}} + m_p}\sigma _{\beta \alpha }q^\alpha \right)D[B]^{*\beta }\gamma_5\bar \Lambda^*_{c[b]}\\
\end{eqnarray}
where $G$, $f$ and $g$ are the strong coupling constants. From this above Lagrangian, the corresponding strong coupling matrix elements can be written as,
\begin{eqnarray}\label{eq:15}
\notag
\left\langle D[B](q)N^*(p') \right|\left. \Lambda _{c[b]}(p) \right\rangle  &&= iG_{\Lambda _{c[b]}D[B]N^*}\bar u(p',s')U(p,s)\\
\notag
\left\langle D[B](q)p(p') \right|\left. \Lambda _{c[b]}(p) \right\rangle  &&= iG_{\Lambda_{c[b]}D[B]p}\bar u(p',s')\gamma _5U(p,s)\\
\notag
\left\langle D[B](q)N^*(p') \right|\left. \Lambda^*_{c[b]}(p) \right\rangle  &&= iG_{\Lambda^*_{c[b]}D[B]N^*}\bar u(p',s')\gamma_5U(p,s)\\
\notag
\left\langle D[B](q)p(p') \right|\left. \Lambda^*_{c[b]}(p) \right\rangle  &&= iG_{\Lambda^*_{c[b]}D[B]p}\bar u(p',s')U(p,s)\\
\notag
\left\langle D^*[B^*](q)N^*(p') \right|\left. \Lambda _{c[b]}(p) \right\rangle  &&= \epsilon^*_\beta\bar u(p',s')\left( f_{\Lambda _{c[b]}D^*[B^*]N^*}\gamma _\beta  \right.\\
\notag
&&\left.  - \frac{ig_{\Lambda _{c[b]}D^*[B^*]N^*}}{m_{\Lambda _{c[b]}} + m_{N^*}}\sigma _{\beta \alpha }q^\alpha \right)\gamma _5U(p,s)\\
\notag
\left\langle D^*[B^*](q)p(p') \right|\left. \Lambda _{c[b]}(p) \right\rangle  &&= \epsilon^*_\beta\bar u(p',s')\left( f_{\Lambda _{c[b]}D^*[B^*]p}\gamma _\beta  \right.\\
\notag
&&\left.  - \frac{ig_{\Lambda _{c[b]}D^*[B^*]p}}{m_{\Lambda _{c[b]}}+ m_p}\sigma _{\beta \alpha }q^\alpha \right)U(p,s)
\end{eqnarray}
\begin{eqnarray}
\notag
\left\langle D^*[B^*](q)N^*(p') \right|\left. \Lambda^*_{c[b]}(p) \right\rangle  &&= \epsilon^*_\beta\bar u(p',s')\left( f_{\Lambda^*_{c[b]}D^*[B^*]N^*}\gamma _\beta \right.\\
\notag
&&\left.  - \frac{ig_{\Lambda^*_{c[b]}D^*[B^*]N^*}}{m_{\Lambda^*_{c[b]}} + m_{N^*}}\sigma _{\beta \alpha }q^\alpha \right)U(p,s)\\
\notag
\left\langle D^*[B^*](q)p(p') \right|\left. \Lambda^*_{c[b]}(p) \right\rangle  &&= \epsilon^*_\beta\bar u(p',s')\left( f_{\Lambda^*_{c[b]}D^*[B^*]p}\gamma _\beta \right.\\
\notag
&&\left.  - \frac{ig_{\Lambda^*_{c[b]}D^*[B^*]p}}{m_{\Lambda^*_{c[b]}} + m_p}\sigma _{\beta \alpha }q^\alpha \right)\gamma_5U(p,s)\\
\end{eqnarray}
where $q=p-p'$, $U(p,s)$ is the spinor wave function of lambda-type charmed (bottom) baryons and $\epsilon_\beta$ is the polarization vector of vector charmed (bottom) mesons.

To derive the sum rules for these strong coupling constants, we firstly introduce the following three-point correlation functions,
\begin{eqnarray}\label{eq:16}
\notag
\Pi (p,p') &&= i^2\int d^4x d^4ye^{ip'x}e^{i(p-p')y} \\
\notag
&&\times \left\langle 0 \right|\mathcal{T}\{ J(x)J_{D[B]}(y){\bar J}_{\Lambda _{c[b]}}(0)\} \left| 0 \right\rangle \\
\notag
\Pi _\mu (p,p') &&= i^2\int d^4x d^4ye^{ip'x}e^{i(p-p')y} \\
&&\times \left\langle 0 \right|\mathcal{T}\{ J(x)J_{D^*[B^*]\mu}(y)\bar J_{\Lambda _{c[b]}}(0)\} \left| 0 \right\rangle
\end{eqnarray}
where $J_{D[B]}(y)$ and $J_{D^*[B^*]\mu}(y)$ are the interpolating currents of pseudoscalar and vector charmed (bottom) mesons and can be expressed as,
\begin{eqnarray}\label{eq:17}
	\notag
	J_{D[B]}(y)&&=\bar{u}(y)i\gamma_5c[b](y)\\ J_{D^*[B^*]\mu}(y)&&=\bar{u}(y)\gamma_\mu c[b](y)
\end{eqnarray}
It is same as the two-point QCD sum rules, the above three-point correlation functions will be handled in both phenomenological and QCD side, respectively.

\subsection{The phenomenological side}\label{sec3.2}
By inserting complete sets of hadron states into the correlation functions, finishing the integral of coordinate space and using the double dispersion relation~\cite{Bracco:2011pg}, the expressions of the three-point correlation function can be written as,
\begin{widetext}
\begin{eqnarray}
\notag
\Pi (p,p') &&= \frac{\left\langle 0 \right|J(0)\left| N^*(p') \right\rangle \left\langle 0 \right|J_{D[B]}(0)\left| D[B](q) \right\rangle \left\langle D[B](q)N^*(p') \right|\left. \Lambda _{c[b]}(p) \right\rangle \left\langle \Lambda _{c[b]}(p) \right|\bar J_{\Lambda _{c[b]}}(0)\left| 0 \right\rangle }{( m_{N^*}^2-p'^2)(m_{\Lambda _{c[b]}}^2-p^2)(m_{D[B]}^2-q^2)}\\
\notag
&&+ \frac{{\left\langle 0 \right|J(0)\left| {p(p')} \right\rangle \left\langle 0 \right|{J_{D[B]}}(0)\left| {D[B](q)} \right\rangle \left\langle {D[B](q)p(p')} \right|\left. {{\Lambda _{c[b]}}(p)} \right\rangle \left\langle {{\Lambda _{c[b]}}(p)} \right|{{\bar J}_{{\Lambda _{c[b]}}}}(0)\left| 0 \right\rangle }}{{( m_p^2-p'^2)(m_{{\Lambda _{c[b]}}}^2-p^2)(m_{D[B]}^2-q^2)}}\\
\notag
&&+ \frac{{\left\langle 0 \right|J(0)\left| {{N^*}(p')} \right\rangle \left\langle 0 \right|{J_{D[B]}}(0)\left| {D[B](q)} \right\rangle \left\langle {D[B](q){N^*}(p')} \right|\left. {\Lambda _{c[b]}^*(p)} \right\rangle \left\langle {\Lambda _{c[b]}^*(p)} \right|{{\bar J}_{{\Lambda _{c[b]}}}}(0)\left| 0 \right\rangle }}{{(m_{{N^*}}^2-p'^2)(m_{\Lambda _{c[b]}^*}^2-p^2)(m_{D[B]}^2-q^2)}}\\
\notag
&&+ \frac{{\left\langle 0 \right|J(0)\left| {p(p')} \right\rangle \left\langle 0 \right|{J_{D[B]}}(0)\left| {D[B](q)} \right\rangle \left\langle {D[B](q)p(p')} \right|\left. {\Lambda _{c[b]}^*(p)} \right\rangle \left\langle {\Lambda _{c[b]}^*(p)} \right|{{\bar J}_{{\Lambda _{c[b]}}}}(0)\left| 0 \right\rangle }}{{(m_p^2-p'^2)(m_{\Lambda _{c[b]}^*}^2-p^2)(m_{D[B]}^2-q^2)}} + ...
\end{eqnarray}
\begin{eqnarray}\label{eq:18}
\notag
{\Pi _\mu }(p,p') &&= \frac{{\left\langle 0 \right|J(0)\left| {{N^*}(p')} \right\rangle \left\langle 0 \right|{J_{{D^*[B^*]}\mu }}(0)\left| {{D^*[B^*]}(q)} \right\rangle \left\langle {{D^*[B^*]}(q){N^*}(p')} \right|\left. {{\Lambda _{c[b]}}(p)} \right\rangle \left\langle {{\Lambda _{c[b]}}(p)} \right|{{\bar J}_{{\Lambda _{c[b]}}}}(0)\left| 0 \right\rangle }}{{(m_{{N^*}}^2-p'^2)(m_{{\Lambda _{c[b]}}}^2-p^2)(m_{D^*[B^*]}^2-q^2)}}\\
\notag
&&+ \frac{{\left\langle 0 \right|J(0)\left| {p(p')} \right\rangle \left\langle 0 \right|{J_{{D^*[B^*]}\mu }}(0)\left| {{D^*[B^*]}(q)} \right\rangle \left\langle {{D^*[B^*]}(q)p(p')} \right|\left. {{\Lambda _{c[b]}}(p)} \right\rangle \left\langle {{\Lambda _{c[b]}}(p)} \right|{{\bar J}_{{\Lambda _{c[b]}}}}(0)\left| 0 \right\rangle }}{{(m_p^2-p'^2)(m_{{\Lambda _{c[b]}}}^2-p^2)(m_{D^*[B^*]}^2-q^2)}}\\
\notag
&&+ \frac{{\left\langle 0 \right|J(0)\left| {{N^*}(p')} \right\rangle \left\langle 0 \right|{J_{{D^*[B^*]}\mu }}(0)\left| {{D^*[B^*]}(q)} \right\rangle \left\langle {{D^*[B^*]}(q){N^*}(p')} \right|\left. {\Lambda _{c[b]}^*(p)} \right\rangle \left\langle {\Lambda _{c[b]}^*(p)} \right|{{\bar J}_{{\Lambda _{c[b]}}}}(0)\left| 0 \right\rangle }}{{(m_{{N^*}}^2-p'^2)(m_{{\Lambda^*_{c[b]}}}^2-p^2)(m_{D^*[B^*]}^2-q^2)}}\\
&&+ \frac{{\left\langle 0 \right|J(0)\left| {p(p')} \right\rangle \left\langle 0 \right|{J_{{D^*[B^*]}\mu }}(0)\left| {{D^*[B^*]}(q)} \right\rangle \left\langle {{D^*[B^*]}(q)p(p')} \right|\left. {\Lambda _{c[b]}^*(p)} \right\rangle \left\langle {\Lambda _{c[b]}^*(p)} \right|{{\bar J}_{{\Lambda _{c[b]}}}}(0)\left| 0 \right\rangle }}{{(m_p^2-p'^2)(m_{\Lambda^*_{c[b]}}^2-p^2)(m_{D^*[B^*]}^2-q^2)}} + ...
\end{eqnarray}
\end{widetext}
The hadron vacuum matrix elements for charmed (bottom) mesons and lambda-type single charmed (bottom) baryons are defined as,
\begin{eqnarray}\label{eq:19}
\notag
\left\langle 0 \right|J_{D[B]}(0)\left| D[B](q) \right\rangle  &&= \frac{f_{D[B]}m_{D[B]}^2}{m_{c[b]}}\\
\notag
\left\langle 0 \right|{J_{{D^*[B^*]}\mu }}(0)\left| {D^*[B^*](q)} \right\rangle  &&= {f_{{D^*[B^*]}}}{m_{{D^*[B^*]}}}{\varepsilon _\mu }\\
\notag
\left\langle 0 \right|{J_{D^*[B^*]\mu }}(0)\left| {{D_0[B_0]}(q)} \right\rangle  &&= {f_{{D_0[B_0]}}}{q_\mu }
\end{eqnarray}
\begin{eqnarray}
\notag
\left\langle {{\Lambda _{c[b]}}(p)} \right|{{\bar J}_{{\Lambda _{c[b]}}}}(0)\left| 0 \right\rangle  &&= {\lambda _{{\Lambda _{c[b]}}}}\bar U(p,s) \\
\left\langle {\Lambda _{c[b]}^*(p)} \right|{{\bar J}_{{\Lambda _{c[b]}}}}(0)\left| 0 \right\rangle  &&=  - {\lambda _{\Lambda _{c[b]}^*}}\bar U(p,s){\gamma _5}
\end{eqnarray}
From these above equations, we can see that the current $J_{\Lambda_{c[b]}}$ (Eq. \ref{eq:1}) also couples to negative parity baryon $\Lambda^*_{c[b]}$. In addition, the vector current $J_{D^*[B^*]\mu}$ can also couple to the scalar charmed (bottom) meson $D_0[B_0]$, its contribution can be eliminated by introducing the projection operator $(g^{\mu\nu}-\frac{q^\mu q^\nu}{q^2})$ in the correlation function.

Substituting the matrix elements in Eq. (\ref{eq:18}) with Eqs. (\ref{eq:5}), (\ref{eq:15}) and (\ref{eq:19}) and multiplied $\Pi_{\mu}(p,p')$ by the projection operator. The correlation functions in phenomenological side can be written as,

\begin{widetext}
\begin{eqnarray}
\notag
	{\Pi ^{\mathrm{phy}}}(p,p') &&= \frac{{{f_{D[B]}}m_{D[B]}^2{\lambda _{{N^*}}}{\lambda _{{\Lambda _{c[b]}}}}{G_{{\Lambda _{c[b]}}D[B]{N^*}}}(\slashed p' + {m_{{N^*}}})(\slashed p + {m_{{\Lambda _{c[b]}}}})}}{{{m_{c[b]}}( m_{{N^*}}^2-p'^2)(m_{{\Lambda _{c[b]}}}^2-p^2)(m_{D[B]}^2-q^2)}} - \frac{{{f_{D[B]}}m_{D[B]}^2{\lambda _p}{\lambda _{{\Lambda _{c[b]}}}}{G_{{\Lambda _{c[b]}}D[B]p}}{\gamma _5}(\slashed p' + {m_p}){\gamma _5}(\slashed p + {m_{{\Lambda _{c[b]}}}})}}{{{m_{c[b]}}(m_p^2-p'^2)(m_{{\Lambda _{c[b]}}}^2-p^2)(m_{D[B]}^2-q^2)}}\\
	\notag
	&&- \frac{{{f_{D[B]}}m_{D[B]}^2{\lambda _{{N^*}}}{\lambda _{\Lambda _c^*}}{G_{\Lambda _{c[b]}^*D[B]{N^*}}}(\slashed p' + {m_{{N^*}}}){\gamma _5}(\slashed p + {m_{\Lambda _{c[b]}^*}}){\gamma _5}}}{{{m_{c[b]}}( m_{{N^*}}^2-p'^2)(m_{\Lambda _{c[b]}^*}^2-p^2)(m_{D[B]}^2-q^2)}} + \frac{{{f_{D[B]}}m_{D[B]}^2{\lambda _p}{\lambda _{\Lambda _{c[b]}^*}}{G_{\Lambda _{c[b]}^*D[B]p}}{\gamma _5}(\slashed p' + {m_p})(\slashed p + {m_{\Lambda _{c[b]}^*}}){\gamma _5}}}{{{m_{c[b]}}(m_p^2-p'^2)(m_{\Lambda _{c[b]}^*}^2-p^2)( m_{D[B]}^2-q^2)}}\\
	\notag
	&&+ ...
\end{eqnarray}
\begin{eqnarray}\label{eq:20}
	\notag
	\tilde \Pi _\nu ^{\mathrm{phy}}(p,p') &&= \left( {{g^{\mu \nu }} - \frac{{{q^\mu }{q^\nu }}}{{{q^2}}}} \right)\Pi _\mu ^{\mathrm{phy}}(p,p') = \left( {{g^{\mu \nu }} - \frac{{{q^\mu }{q^\nu }}}{{{q^2}}}} \right)\\
	\notag
	&&\times \left[ {\frac{{{\lambda _{{N^*}}}{\lambda _{{\Lambda _{c[b]}}}}{f_{{D^*[B^*]}}}{m_{{D^*[B^*]}}}\left( -{{g_{\mu \beta }} + \frac{{{q^\mu }{q^\beta }}}{{{m_{D^*[B^*]}^2}}}} \right)(\slashed p' + {m_{{N^*}}})\left( {{f_{{\Lambda _{c[b]}}{D^*[B^*]}{N^*}}}{\gamma _\beta } - i\frac{{{g_{{\Lambda _{c[b]}}{D^*[B^*]}{N^*}}}}}{{{m_{{\Lambda _{c[b]}}}} + {m_{{N^*}}}}}{\sigma ^{\beta \alpha }}{q_\alpha }} \right){\gamma _5}(\slashed p + {m_{{\Lambda _{c[b]}}}})}}{{(m_{{N^*}}^2-p'^2)(m_{{\Lambda _{c[b]}}}^2-p^2)( m_{D^*[B^*]}^2-q^2)}}} \right.\\
	\notag
	&&- \frac{{{\lambda _p}{\lambda _{{\Lambda _{c[b]}}}}{f_{{D^*[B^*]}}}{m_{{D^*[B^*]}}}\left( -{{g_{\mu \beta }} + \frac{{{q^\mu }{q^\beta }}}{{{m_{D^*[B^*]}^2}}}} \right){\gamma _5}(\slashed p' + {m_p})\left( {{f_{{\Lambda _{c[b]}}{D^*[B^*]}p}}{\gamma _\beta } - i\frac{{{g_{{\Lambda _{c[b]}}{D^*[B^*]}p}}}}{{{m_{{\Lambda _{c[b]}}}} + {m_p}}}{\sigma ^{\beta \alpha }}{q_\alpha }} \right)(\slashed p + {m_{{\Lambda _{c[b]}}}})}}{{(m_p^2-p'^2)(m_{{\Lambda _{c[b]}}}^2-p^2)(m_{D^*[B^*]}^2-q^2)}}\\
	\notag
	&&- \frac{{{\lambda _{{N^*}}}{\lambda _{\Lambda _{c[b]}^*}}{f_{{D^*[B^*]}}}{m_{{D^*[B^*]}}}\left( -{{g_{\mu \beta }} + \frac{{{q^\mu }{q^\beta }}}{{{m_{D^*[B^*]}^2}}}} \right)(\slashed p' + {m_{{N^*}}})\left( {{f_{\Lambda _{c[b]}^*{D^*[B^*]}{N^*}}}{\gamma _\beta } - i\frac{{{g_{\Lambda _{c[b]}^*{D^*[B^*]}{N^*}}}}}{{{m_{\Lambda _{c[b]}^*}} + {m_{{N^*}}}}}{\sigma ^{\beta \alpha }}{q_\alpha }} \right)(\slashed p + {m_{\Lambda _{c[b]}^*}}){\gamma _5}}}{{(m_{{N^*}}^2-p'^2)(m_{\Lambda _{c[b]}^*}^2-p^2)(m_{{D^*[B^*]}}^2-q^2)}}\\
	\notag
	&&\left. { + \frac{{{\lambda _p}{\lambda _{\Lambda _{c[b]}^*}}{f_{{D^*[B^*]}}}{m_{{D^*[B^*]}}}\left( -{{g_{\mu \beta }} + \frac{{{q^\mu }{q^\beta }}}{{{m_{D^*[B^*]}^2}}}} \right){\gamma _5}(\slashed p' + {m_p})\left( {{f_{\Lambda _{c[b]}^*{D^*[B^*]}p}}{\gamma _\beta } - i\frac{{{g_{\Lambda _{c[b]}^*{D^*[B^*]}p}}}}{{{m_{\Lambda _{c[b]}^*}} + {m_p}}}{\sigma ^{\beta \alpha }}{q_\alpha }} \right){\gamma _5}(\slashed p + {m_{\Lambda _{c[b]}^*}}){\gamma _5}}}{{(m_p^2-p'^2)(m_{\Lambda _{c[b]}^*}^2-p^2)(m_{{D^*[B^*]}}^2-q^2)}} + ...} \right]\\
\end{eqnarray}
\end{widetext}
Finally, these above correlation functions can be decomposed into different dirac structures,
\begin{eqnarray}\label{eq:21}
\notag
\Pi ^{\mathrm{phy}}(p,p') &&= \Pi _1^{\mathrm{phy}}\slashed p' + \Pi _2^{\mathrm{phy}}\slashed p + \Pi _3^{\mathrm{phy}}\slashed p'\slashed p + \Pi _4^{\mathrm{phy}}\\
\notag
\tilde \Pi _\nu ^{\mathrm{phy}}(p,p') &&=\tilde \Pi _1^{\mathrm{phy}}\gamma _\nu \gamma _5 + \tilde \Pi _2^{\mathrm{phy}}\gamma _\nu \slashed p'\gamma _5 +\tilde \Pi _3^{\mathrm{phy}}\gamma _\nu \slashed q\gamma _5  \\
\notag
&&+ \tilde \Pi _4^{\mathrm{phy}}\gamma _\nu \slashed p' \slashed q \gamma _5+ \tilde \Pi _5^{\mathrm{phy}}\slashed p'\gamma _5 p'_\nu + \tilde \Pi _6^{\mathrm{phy}}\slashed p'\gamma _5 q_\nu \\
&&+ \tilde \Pi _7^{\mathrm{phy}}\slashed q\gamma _5 p'_\nu + \tilde \Pi _8^{\mathrm{phy}}\slashed q\gamma _5 q_\nu+...
\end{eqnarray}
The strong coupling constants are included in these expansion coefficients $\Pi^{\mathrm{phy}}_i$ and $\tilde{\Pi}^{\mathrm{phy}}_i$ which are commonly named as scalar invariant amplitudes. It is worth noting that when we consider the contribution of all possible positive and negative parity baryons, the correlation functions for vertex $\Lambda_{c[b]}D[B]N^*$ contains four strong coupling constants, and eight strong coupling constants for vertex $\Lambda_{c[b]}D^*[B^*]N^*$. In order to determine the value of these coupling constants, four and eight independent scalar invariant amplitudes in Eq. (\ref{eq:21}) must be extracted to achieve this goal.

\subsection{The QCD side}\label{sec3.3}
\begin{figure*}
\centering
\includegraphics[width=16cm]{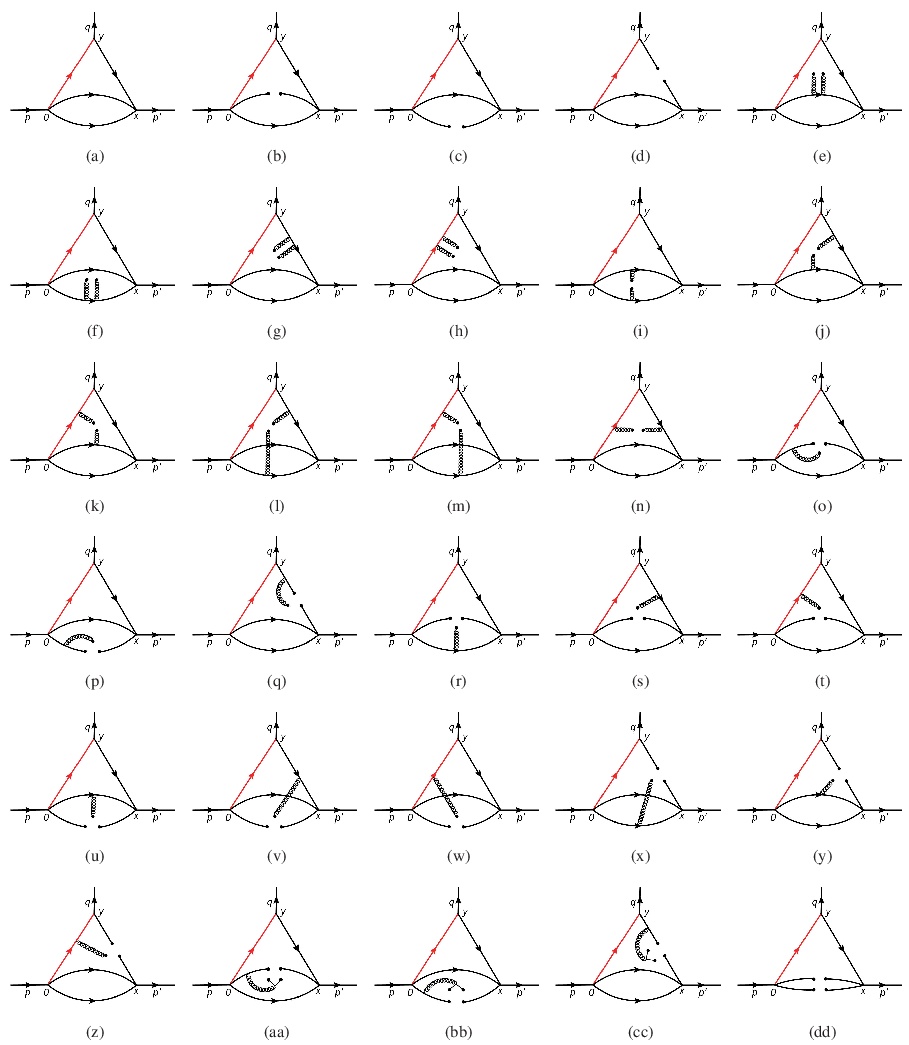}
\caption{The Feynman diagrams for the perturbative part and vacuum condensate terms, where the black and red solid lines denote the $u$ or $d$ and $c$ or $b$ quark lines, respectively. The black loop lines are the gluon lines.}
\label{FD}
\end{figure*}

In QCD sides, we bring the specific form of interpolating currents into the three-point correlation functions in Eq. (\ref{eq:16}). Then, the correlation functions in QCD side can be expressed as the following forms after doing the OPE,
\begin{eqnarray}\label{eq:22}
\notag
\Pi ^{\mathrm{QCD}}(p,p') &&=  - 2i\varepsilon _{i'j'k'}\varepsilon _{ijk}\int d^4xd^4ye^{ip'x}e^{i(p-p')y}\{ \gamma _\beta D^{k'j}(x) \\
\notag
&& \times \gamma _5\mathcal{C}U^{ki'T}(x)\mathcal{C}\gamma _\beta U^{j'm}(x - y)\gamma _5C[B]^{mi}(y)\} \\
\notag
\Pi _\mu ^{\mathrm{QCD}}(p,p') &&=  - 2\varepsilon _{i'j'k'}\varepsilon _{ijk}\int d^4x d^4ye^{ip'x}e^{i(p-p')y} \{ \gamma _\beta D^{k'j}(x) \\
\notag
&& \times \gamma _5\mathcal{C}U^{ki'T}(x)\mathcal{C}\gamma _\beta U^{j'm}(x - y)\gamma _\mu C[B]^{mi}(y) \} \\
\end{eqnarray}
where $C[B]^{ij}(x)$ is the full propagator of $c[b]$ quark and can be written as the following form in momentum space~\cite{Pascual:1984zb,Reinders:1984sr},
\begin{eqnarray}\label{eq:23}
\notag
C[B]^{ij}(x) &&= \frac{i}{(2\pi )^4}\int d^4k e^{- ikx}\left\{ \frac{\delta ^{ij}}{\slashed k - m_{c[b]}} \right.\\
\notag
&& - \frac{{{g_s}G_{\alpha \beta }^nt_{ab}^n}}{4}\frac{{{\sigma ^{\alpha \beta }}(\slashed{k} + {m_{c[b]}}) + (\slashed{k} + {m_{c[b]}}){\sigma ^{\alpha \beta }}}}{{{{({k^2} - m_{c[b]}^2)}^2}}}\\
\notag
&& - \frac{{g_s^2{{({t^m}{t^n})}_{ab}}G_{\alpha \beta }^mG_{\mu \nu }^n({f^{\alpha \beta \mu \nu }} + {f^{\alpha \mu \beta \nu }} + {f^{\alpha \mu \nu \beta }})}}{{4{{({k^2} - m_{c[b]}^2)}^5}}}\\
&&\left. { + ...} \right\}
\end{eqnarray}
Here, $f^{\lambda\alpha\beta}$ and $f^{\alpha\beta\mu\nu}$ have the following forms,
\begin{eqnarray}
\notag
{f^{\lambda \alpha \beta }} &&= (\slashed k + {m_{c[b]}}){\gamma ^\lambda }(\slashed k + {m_{c[b]}}){\gamma ^\alpha }(\slashed k + {m_{c[b]}}){\gamma ^\beta }(\slashed k + {m_{c[b]}})\\
\notag
{f^{\alpha \beta \mu \nu }} &&= (\slashed k + {m_{c[b]}}){\gamma ^\alpha }(\slashed k + {m_{c[b]}}){\gamma ^\beta }(\slashed k + {m_{c[b]}})\\
&& \times {\gamma ^\mu }(\slashed k + {m_{c[b]}}){\gamma ^\nu }(\slashed k + {m_{c[b]}})
\end{eqnarray}
By using the double dispersion relation, the correlation functions in QCD side can be written as,
\begin{eqnarray}\label{eq:25}
\notag
\Pi ^{\mathrm{QCD}}(p,p') &&= \int\limits_{s_{\min}}^\infty  {ds} \int\limits_{u_{\min}}^\infty  {du} \frac{\rho ^{\mathrm{QCD}}(s,u,q^2)}{{(s - p^2)(u - p'^2)}}\\
\Pi _\mu ^{\mathrm{QCD}}(p,p') &&= \int\limits_{s_{\min}}^\infty  {ds} \int\limits_{u_{\min}}^\infty  {du} \frac{{\rho _\mu ^{\mathrm{QCD}}(s,u,q^2)}}{{(s - p^2)(u - p'^2)}}
\end{eqnarray}
where $\rho^{\mathrm{QCD}}(s,u,q^2)$ and $\rho^{\mathrm{QCD}}_{\mu}(s,u,q^2)$ are QCD spectral density with $s=p^2$ and $u=p'^2$. $s_{\min}$ and $u_{\min}$ are create thresholds for $\Lambda_{c[b]}$ and $N^*$ which are taken as $m_{c[b]}^2$ and $0$, respectively.
The correlation functions in QCD side can also be written as,
\begin{eqnarray}\label{eq:26}
\notag
\Pi ^{\mathrm{QCD}}(p,p') &&= \Pi _1^{\mathrm{QCD}}\slashed p' + \Pi _2^{\mathrm{QCD}}\slashed p + \Pi _3^{\mathrm{QCD}}\slashed p'\slashed p + \Pi _4^{\mathrm{QCD}}\\
\notag
\tilde \Pi _\nu ^{\mathrm{QCD}}(p,p') &&=\tilde \Pi _1^{\mathrm{QCD}}\gamma _\nu \gamma _5 + \tilde \Pi _2^{\mathrm{QCD}}\gamma _\nu \slashed p'\gamma _5 +\tilde \Pi _3^{\mathrm{QCD}}\gamma _\nu \slashed q\gamma _5  \\
\notag
&&+ \tilde \Pi _4^{\mathrm{QCD}}\gamma _\nu \slashed p' \slashed q \gamma _5+ \tilde \Pi _5^{\mathrm{QCD}}\slashed p'\gamma _5 p'_\nu + \tilde \Pi _6^{\mathrm{QCD}}\slashed p'\gamma _5 q_\nu \\
&&+ \tilde \Pi _7^{\mathrm{QCD}}\slashed q\gamma _5 p'_\nu + \tilde \Pi _8^{\mathrm{QCD}}\slashed q\gamma _5 q_\nu+...
\end{eqnarray}
where $\Pi^{\mathrm{QCD}}_i$ and $\tilde \Pi^{\mathrm{QCD}}_i$ are scalar invariant amplitudes in QCD side, and can be represented as the summation of perturbative part and vacuum condensate terms. The Feynman diagrams about the perturbative part, vacuum condensate terms $\langle{\bar qq}\rangle$, $\langle g_{s}^{2}GG\rangle$, $\langle \bar q g_{s}\sigma Gq\rangle$, $\langle{\bar qq}\rangle^{2}$ and $g_{s}^{2}\langle{\bar qq}\rangle^{2}$ are explicitly shown in Fig. \ref{FD}.

For the contribution of perturbative term, we substitude the free propagators in momentum space of both light and heavy quarks in Eq. (\ref{eq:22}). The corresponding Feynman diagram is shown as Fig. \ref{FD} (a), and its contributions are given as,
\begin{eqnarray}\label{eq:27}
\notag
{\Pi ^{\mathrm{QCD}}_0}(p,p') &&=  - \frac{{12i}}{{{{(2\pi )}^8}}}\int {{d^4}{k_1}} {d^4}{k_2}{d^4}{k_3}{d^4}{k_4}{\delta ^4}(q + {k_3} - {k_4})\\
\notag
&& \times {\delta ^4}(p' - {k_1} - {k_2} - {k_3}) \\
\notag
&&\times \frac{{{\gamma _\beta }{{\slashed k}_1}{\gamma _5}{{\slashed k}_2}{\gamma _\beta }{{\slashed k}_3}{\gamma _5}({{\slashed k}_4} + {m_{c[b]}})}}{{k_1^2k_2^2k_3^2(k_4^2 - m_{c[b]}^2)}}\\
\notag
\Pi _{0\mu}^{\mathrm{QCD}}(p,p') &&=  - \frac{12}{{{{(2\pi )}^8}}}\int {{d^4}{k_1}} {d^4}{k_2}{d^4}{k_3}{d^4}{k_4}{\delta ^4}(q + {k_3} - {k_4})\\
\notag
&& \times {\delta ^4}(p' - {k_1} - {k_2} - {k_3}) \\
&&\frac{{{\gamma _\beta}{{\slashed k}_1}{\gamma _5}{{\slashed k}_2}{\gamma _\beta }{{\slashed k}_3}{\gamma _\mu}({{\slashed k}_4} + {m_{c[b]}})}}{{k_1^2k_2^2k_3^2(k_4^2 - m_{c[b]}^2)}}
\end{eqnarray}
Here, the subscripts zero denotes the dimension of perturbative term. For simplicity, we ignore the mass of $u(d)$ quark in the following calculations. By setting all quark lines on-shell with the Cutkosky's rule~\cite{Cutkosky:1960sp}, the QCD spectral density functions can be obtained,
\begin{eqnarray}\label{eq:28}
\notag
{\rho ^{\mathrm{QCD}}_0}(s,u,q^2) &&= \frac{{12i}}{{{{(2\pi )}^8}}}\frac{{{{( - 2\pi i)}^5}}}{{{{(2\pi i)}^3}}}\int\limits_0^u {d\xi } \int {{d^4}{k_1}} \delta [{(q' - {k_1})^2}]\\
\notag
&& \times \delta (k_1^2) \int {{d^4}{k_3}} \delta [{(p' - {k_3})^2} - \xi ]\delta (k_3^2)\\
\notag
&& \times \delta [{({k_3} + q)^2} - m_{c[b]}^2] {\gamma _\beta }{{\slashed k}_1}{\gamma _5}(\slashed p' - {{\slashed k}_3} - {{\slashed k}_1}) \\
\notag
&& \times{\gamma _\beta }{{\slashed k}_3}{\gamma _5}(\slashed q + {{\slashed k}_3} + {m_{c[b]}})\\
\notag
\rho _{0\mu} ^{\mathrm{QCD}}(s,u,{q^2}) &&= \frac{{12}}{{{{(2\pi )}^8}}}\frac{{{{( - 2\pi i)}^5}}}{{{{(2\pi i)}^3}}}\int\limits_0^u {d\xi } \int {{d^4}{k_1}}\delta [{(q' - {k_1})^2}]\\
\notag
&&\times \delta (k_1^2) \int {{d^4}{k_3}} \delta [{(p' - {k_3})^2} - \xi ]\delta (k_3^2)\\
\notag
&& \times \delta [{({k_3} + q)^2} - m_{c[b]}^2] {\gamma _\beta}{{\slashed k}_1}{\gamma _5}(\slashed p' - {{\slashed k}_3} - {{\slashed k}_1})\\
&&\times{\gamma _\beta }{{\slashed k}_3}{\gamma _\mu}(\slashed q + {{\slashed k}_3} + {m_{c[b]}})
\end{eqnarray}
where $q'=k_{1}+k_{2}$ and $\xi=q'^{2}$. The integral formulas for two and three Dirac delta functions are shown in Appendix. \ref{Sec:AppA}.

The contribution of quark condensate $\langle\bar qq\rangle$ comes from the second term of the light quark full propagator in Eq. (\ref{eq:9}). The corresponding Feynman diagrams are shown in Figs. \ref{FD} (b) $\sim$ (d). Since the heavy quark will not contribute to this condensation, there are three Feynman diagrams from light quark condensate. For contribution in Fig. \ref{FD} (b) as an example, it can be expressed as,
\begin{eqnarray}\label{eq:29}
\notag
\Pi _{3b}^{\mathrm{QCD}}(p,p') &&= \frac{\left\langle {\bar qq} \right\rangle i}{{{{(2\pi )}^4}}}\int {{d^4}{k_3}} \\
\notag
&& \times \frac{{{\gamma _\beta }{\gamma _5}(\slashed p' - {{\slashed k}_3}){\gamma _\beta }{{\slashed k}_3}{\gamma _5}({{\slashed k}_3} + \slashed q + {m_{c[b]}})}}{{k_3^2{{(p' - {k_3})}^2}[{{(q + {k_3})}^2} - m_{c[b]}^2]}}\\
\notag
\Pi _{3b\mu }^{\mathrm{QCD}}(p,p') &&= \frac{{\left\langle {\bar qq} \right\rangle }}{{{{(2\pi )}^4}}}\int {{d^4}{k_3}} \\
\notag
&& \times \frac{{{\gamma _\beta}{\gamma _5}(\slashed p' - {{\slashed k}_3}){\gamma _5 }{{\slashed k}_3}{\gamma _\mu}({{\slashed k}_3} + \slashed q + {m_{c[b]}})}}{{k_3^2{{(p' - {k_3})}^2}[{{(q + {k_3})}^2} - m_{c[b]}^2]}}\\
\end{eqnarray}
The corresponding QCD spectral density functions can also be obtained by Cutkosky's rule,
\begin{eqnarray}\label{eq:30}
\notag
\rho _{3b}^{\mathrm{QCD}}(s,u,q^2) &&= \frac{\left\langle {\bar qq} \right\rangle i}{{{{(2\pi )}^4}}}\frac{{{{( - 2\pi i)}^3}}}{{{{(2\pi i)}^2}}}\int {{d^4}{k_3}} \delta (k_3^2)\delta [{(p' - {k_3})^2}] \\
\notag
&& \times \delta [{(q + {k_3})^2} - m_{c[b]}^2]{\gamma _\beta }{\gamma _5}(\slashed p' - {{\slashed k}_3})\\
\notag
&& \times{\gamma _\beta }{{\slashed k}_3}{\gamma _5}({{\slashed k}_3} + \slashed q + {m_{c[b]}})
\end{eqnarray}
\begin{eqnarray}
\notag
\rho _{3b\mu }^{\mathrm{QCD}}(s,u,q^2) &&= \frac{{\left\langle {\bar qq} \right\rangle }}{{{{(2\pi )}^4}}}\frac{{{{( - 2\pi i)}^3}}}{{{{(2\pi i)}^2}}}\times \int {{d^4}{k_3}} \delta (k_3^2)\delta [{(p' - {k_3})^2}]\\
\notag
&& \delta [{(q + {k_3})^2} - m_{c[b]}^2]{\gamma _\beta}{\gamma _5}(\slashed p' - {{\slashed k}_3})\\
&& \times{\gamma _5 }{{\slashed k}_3}{\gamma _\mu}({{\slashed k}_3} + \slashed q + {m_{c[b]}})
\end{eqnarray}

For the gluon condensate $\langle g_{s}^{2}GG\rangle$, there are ten Feynman diagrams which are shown in Figs. \ref{FD} (e) $\sim$ (n). For Fig. \ref{FD} (h) as an example, this contribution can be written as,
\begin{eqnarray}\label{eq:30}
\notag
\Pi _{4h}^{\mathrm{QCD}}(p,p') &&= \frac{{\left\langle {g_s^2GG} \right\rangle i}}{{24{{(2\pi )}^8}}}\int {{d^4}{k_1}{d^4}{k_2}{d^4}{k_3}{d^4}{k_4}}\\
\notag
&&\times \delta^{4} (q + {k_3} - {k_4})\delta^{4} (p' - {k_1} - {k_2} - {k_3}) \\
\notag
&& \times ({g_{\xi \lambda }}{g_{\zeta \chi }} - {g_{\xi \chi }}{g_{\zeta \lambda }})\\
\notag
&& \times \frac{{{\gamma _\beta }{{\slashed k}_1}{\gamma _5}{{\slashed k}_2}{\gamma _\beta }{{\slashed k}_3}{\gamma _5}({f^{\lambda \chi \xi \zeta }} + {f^{\lambda \xi \chi \zeta }} + {f^{\lambda \xi \zeta \chi }})}}{{k_1^2k_2^2k_3^2{{(k_4^2 - m_{c[b]}^2)}^5}}}
\end{eqnarray}
\begin{eqnarray}
\notag
\Pi _{4h\mu }^{\mathrm{QCD}}(p,p') &&= \frac{{\left\langle {g_s^2GG} \right\rangle }}{{24{{(2\pi )}^8}}}\int {{d^4}{k_1}{d^4}{k_2}{d^4}{k_3}{d^4}{k_4}} \\
\notag
&& \times \delta^{4}(q + {k_3} - {k_4}) \delta^{4}(p' - {k_1} - {k_2} - {k_3})\\
\notag
&&\times ({g_{\xi \lambda }}{g_{\zeta \chi }} - {g_{\xi \chi }}{g_{\zeta \lambda }}) \\
\notag
&& \times \frac{{{\gamma _\beta}{{\slashed k}_1}{\gamma _5}{{\slashed k}_2}{\gamma _5 }{{\slashed k}_3}{\gamma _\mu}({f^{\lambda \chi \xi \zeta }} + {f^{\lambda \xi \chi \zeta }} + {f^{\lambda \xi \zeta \chi }})}}{{k_1^2k_2^2k_3^2{{(k_4^2 - m_{c[b]}^2)}^5}}}\\
\end{eqnarray}
In Eq. (\ref{eq:31}), $1/(k_4^2-m_{c[b]}^2)^{5}$ can be handled by the following formula,
\begin{eqnarray}\label{eq:31}
\frac{1}{{{{({k^2} - m_{c[b]}^2)}^n}}} = {\left. {\frac{1}{{(n - 1)!}}\frac{{{\partial ^n}}}{{{{(\partial A)}^n}}}\frac{1}{{{k^2} - A}}} \right|_{A \to m_{c[b]}^2}}
\end{eqnarray}
Again, the spectral density function can be derived by the Cutkosky's rule after using the above derivative formula.

For quark gluon mixed condensate $\langle\bar{q}g_{s}\sigma Gq\rangle$ which are shown in Figs. \ref{FD} (o)$\sim$(z), there are two mechanisms to produce this term. The first is that a light quark line emits a gluon and a quark-antiquark pair to generate the mixed condensation (Figs. \ref{FD} (o)$\sim$(q)). This contribution originates from the third term in light quark full propagator in Eq. (\ref{eq:8}). Taking Fig. \ref{FD} (o) as an example, by using variable replacement $x_{\mu}\to
i\partial /\partial p{'_\mu }$, this contribution can be written as,
\begin{eqnarray}\label{eq:32}
\notag
\Pi _{5o}^{\mathrm{QCD}}(p,p') &&=  - \frac{\left\langle {\bar q{g_s}\sigma Gq} \right\rangle i}{{16{{(2\pi )}^4}}}{({\partial _{p'}})^2}\int {{d^4}{k_3}} \\
\notag
&& \times \frac{{{\gamma _\beta }{\gamma _5}(\slashed p' - {{\slashed k}_3}){\gamma _\beta }{{\slashed k}_3}{\gamma _5}({{\slashed k}_3} + \slashed q + {m_{c[b]}})}}{{k_3^2{{(p' - {k_3})}^2}[{{({k_3} + q)}^2} - m_{c[b]}^2]}}
\end{eqnarray}
\begin{eqnarray}
\notag
\Pi _{5o\mu }^{\mathrm{QCD}}(p,p') &&=  - \frac{{\left\langle {\bar q{g_s}\sigma Gq} \right\rangle }}{{16{{(2\pi )}^4}}}{({\partial _{p'}})^2}\int {{d^4}{k_3}} \\
\notag
&& \times \frac{{{\gamma _
		\beta}{\gamma _5}(\slashed p' - {{\slashed k}_3}){\gamma _5 }{{\slashed k}_3}{\gamma _\mu}({{\slashed k}_3} + \slashed q + {m_{c[b]}})}}{{k_3^2{{(p' - {k_3})}^2}[{{({k_3} + q)}^2} - m_{c[b]}^2]}} \\
\end{eqnarray}
It is the similar as gluon condensate, the QCD spectral density can also be obtained according to Cutkosky's rule. The second mechanism to produce this mixed condensate is that one light or heavy quark line emits gluon and another light quark line contributes a quark-antiquark pair to generate this condensation (Figs. \ref{FD} (r)$\sim$(z)). This kind of contribution comes from the second term in Eq. (\ref{eq:23}) and the fourth term in Eq. (\ref{eq:9}). The denominator power in the factor $1/(k^{2}-m^{2})^{n}$ that will appear in Feynman integral can also be reduced by using derivative formula in Eq. (\ref{eq:32}).

The contributions of four quark condensations are shown in Figs. \ref{FD} (aa)$\sim$(dd). Figs. \ref{FD} (aa)$\sim$(cc) describe that a light quark line contributes a quark-antiquark pair and a gluon, and then the gluon contributes a quark-antiquark pair to product the four quark condensation. This contribution comes from the sixth term in Eq. (\ref{eq:9}) and can be calculated by doing variable replacement $x_{\mu}\to i\partial /\partial p{'_\mu }$. Fig. \ref{FD} (dd) describes that each of two light quark lines contribute a quark-antiquark pair to produce four quark condensation. This contribution plays a key role to the final results and can be expressed as the following form,
\begin{eqnarray}\label{eq:34}
\notag
\Pi _{6dd}^{\mathrm{QCD}}(p,p') &&=  - \frac{{i{{\left\langle {\bar qq} \right\rangle }^2}}}{{12}}\frac{{{\gamma _\beta }{\gamma _5}{\gamma _\beta }\slashed p'{\gamma _5}(\slashed p'+\slashed q + m_{c[b]})}}{{p{'^2}({p^2} - m_{c[b]}^2)}}\\
\Pi _{6dd\mu }^{\mathrm{QCD}}(p,p') &&=  - \frac{{{{\left\langle {\bar qq} \right\rangle }^2}}}{{12}}\frac{{{\gamma _\beta}{\gamma _5}{\gamma _\beta }\slashed p'{\gamma _\mu}(\slashed p'+\slashed q + {m_{c[b]}})}}{{p{'^2}({p^2} - m_{c[b]}^2)}}
\end{eqnarray}

\subsection{The strong coupling constants}\label{sec3.4}

By taking the change of variables $p^2\to-P^2$, $p'^2\to-P'^2$ and $q^2\to-Q^2$, we can perform double Borel transformation for variables $P^2$ and $P'^2$ to both phenomenological and QCD sides which can further suppress the contribution of higher resonance and continuum states in phenomenological side. Furthermore, it also suppresses the higher dimension operators in QCD side and improves the convergence of OPE~\cite{Reinders:1984sr}. After double Borel transformation, the variables $P^2$ and $P'^2$ will be replaced by Borel parameters $T_1^2$ and $T_2^2$ which can be reduced as one parameter with the relations $T^2=T_1^2$ and $T_2^2=kT_1^2=kT^2$ ($k=\frac{m^2_{N^*}}{m^2_{\Lambda_{c[b]}}-m^2_{c[b]}}$)~\cite{Zhang:2023nxl,Leljak:2021vte}. Finally, using quark-hadron duality condition, we can establish a series of linear equations about different scalar invariant amplitudes. By solving these linear equations, the momentum dependent strong coupling constants $G_{\Lambda_{c[b]}D[B]N^*}(Q^2)$, $f_{\Lambda_{c[b]}D^*[B^*]N^*}(Q^2)$ and $g_{\Lambda_{c[b]}D^*[B^*]N^*}(Q^2)$ can be expressed as,
\begin{widetext}
\begin{eqnarray}\label{eq:35}
\notag
G_{\Lambda _{c[b]}D[B]N^*}(Q^2) &&= \frac{{{m_{c[b]}}(Q^2 + m_{D[B]}^2)}}{{{f_{D[B]}}m_{D[B]}^2{\lambda _{{\Lambda _{c[b]}}}}{\lambda _{{N^*}}}}}{e^{m_{{\Lambda _{c[b]}}}^2/T^2 + m_{{N^*}}^2/kT^2}}\left\{ {\int\limits_{{s_{\min }}}^{{s_0}} {ds} \int\limits_{{u_{\min }}}^{{u_0}} {du\hat \rho (s,u,{Q^2}){e^{ - s/T^2 - u/kT^2}}}  + \mathcal{BB}[{{\hat \Pi }^{{{\left\langle {\bar qq} \right\rangle }^2}}}]} \right\}\\
\notag
{f_{{\Lambda _{c[b]}}{D^*[B^*]}{N^*}}}(Q^2) &&= \frac{{(Q^2 + m_{D^*[B^*]}^2)}}{{{f_{{D^*[B^*]}}}{m_{{D^*[B^*]}}}{\lambda _{{\Lambda _{c[b]}}}}{\lambda _{{N^*}}}}}{e^{m_{{\Lambda _{c[b]}}}^2/T^2 + m_{{N^*}}^2/kT^2}}\left\{ {\int\limits_{{s_{\min }}}^{{s_0}} {ds} \int\limits_{{u_{\min }}}^{{u_0}} {du{{\bar \rho }_1}(s,u,{Q^2}){e^{ - s/T^2 - u/kT^2}}}  + \mathcal{BB}[{{\bar \Pi }_1}^{{{\left\langle {\bar qq} \right\rangle }^2}}]} \right\}\\
{g_{{\Lambda _{c[b]}}{D^*[B^*]}{N^*}}}({Q^2}) &&= \frac{{({Q^2} + m_{{D^*[B^*]}}^2)}}{{{f_{{D^*[B^*]}}}{m_{{D^*[B^*]}}}{\lambda _{{\Lambda _{c[b]}}}}{\lambda _{{N^*}}}}}{e^{m_{{\Lambda _{c[b]}}}^2/T^2 + m_{{N^*}}^2/kT^2}}\left\{ {\int\limits_{{s_{\min }}}^{{s_0}} {ds} \int\limits_{{u_{\min }}}^{{u_0}} {du{{\bar \rho }_2}(s,u,{Q^2}){e^{- s/T^2 - u/kT^2}}}  + \mathcal{BB}[{{\bar \Pi }_2}^{{{\left\langle {\bar qq} \right\rangle }^2}}]} \right\}
\end{eqnarray}
\end{widetext}
Here, $\mathcal{BB}$ denotes the double Borel transformation to four quark condensation, $s_{0}$ and $u_{0}$ are threshold parameters for initial and final state hadrons, respectively. The QCD spectral density functions $\hat\rho(s,u,q^2)$ and $\bar\rho_{1,2}(s,u,Q^2)$ include contributions of perturbative part and different condensate terms($\rho_i(s,u,Q^2)$ and $\tilde\rho_i(s,u,Q^2)$), and also contain contributions of different dirac structures. The full expressions of these functions are too complex to be shown here for simplicity.

\section{Numerical results and Discussions}\label{sec4}
The input parameters in the present work are all listed in Table \ref{IP},
\begin{table}[htbp]
	\begin{ruledtabular}\caption{Input parameters (IP) in this analysis. The values of vacuum condensate are at the energy scale $\mu=1$ GeV.}
		\label{IP}
		\begin{tabular}{c c c c}
			IP&Values(GeV)~\cite{ParticleDataGroup:2024cfk}&IP&Values\\ \hline
			$m_{\Lambda_{c}}$&2.286&$f_D$&$0.210\pm0.011$ GeV~\cite{Wang:2015mxa}\\
			$m_{\Lambda^*_{c}}$&2.592&$f_{D^*}$&$0.236\pm0.021$ GeV~\cite{Wang:2015mxa}\\
			$m_{\Lambda_{b}}$&5.619&$f_B$&$0.192\pm0.013$ GeV~\cite{Wang:2015mxa}\\
			$m_{\Lambda^*_{b}}$&5.912&$f_{B^*}$&$0.213\pm0.018$ GeV~\cite{Wang:2015mxa}\\
			$m_{D}$&1.86&$\lambda_{\Lambda_c}$&$0.0151\pm0.0023$ GeV$^{3}$~\cite{Wang:2020mxk}\\
			$m_{D^*}$&2.01&$\lambda_{\Lambda_b}$&$0.0196\pm0.0036$ GeV$^{3}$~\cite{Wang:2020mxk}\\
			$m_{B}$&5.28&$\langle\overline{q}q\rangle$&$-(0.23\pm0.01)^{3}$ GeV$^{3}$~\cite{Shifman:1978by,Reinders:1984sr}\\
			$m_{B^*}$&5.32&$\langle\overline{q}g_{s}\sigma Gq\rangle$&$m_{0}^{2}\langle\overline{q}q\rangle$~\cite{Shifman:1978by,Reinders:1984sr}\\
			$m_{N^*}$&1.535&$m_{0}^{2}$&$0.8\pm0.1$ GeV$^{2}$~\cite{Shifman:1978by,Reinders:1984sr}\\
			$m_p$&0.983&$\langle g_{s}^{2}GG\rangle$&$0.47\pm0.15$ GeV$^{4}$~\cite{Narison:2010cg,Narison:2011xe,Narison:2011rn}
		\end{tabular}
	\end{ruledtabular}
\end{table}
The masses of $c$ and $b$ quarks and the values of vacuum condensates are energy dependent, which can be expressed as follows by using the renormlization group equation (RGE),
\begin{eqnarray}\label{eq:36}
\notag
m_{c[b]}(\mu ) &&= m_{c[b]}(m_{c[b]})\left[\frac{\alpha _s(\mu )}{\alpha _s(m_{c[b]})}\right]^{\frac{12}{33 - 2n_f}}\\
\notag
\left\langle \bar qq \right\rangle (\mu ) &&= \left\langle \bar qq \right\rangle (1\mathrm{GeV})\left[\frac{\alpha _s(1\mathrm{GeV})}{\alpha _s(\mu )}\right]^{\frac{12}{33 - 2n_f}}\\
\notag
\left\langle \bar qg_s\sigma Gq \right\rangle (\mu ) &&= \left\langle \bar qg_s\sigma Gq \right\rangle (1\mathrm{GeV})\left[\frac{\alpha _s(1\mathrm{GeV})}{\alpha _s(\mu )}\right]^{\frac{2}{33 - 2n_f}}
\end{eqnarray}
\begin{eqnarray}
\notag
\alpha _s(\mu ) &&= \frac{1}{b_0t}\left[ 1 - \frac{b_1}{b_0^2}\frac{\log t}{t} \right.\\
&&\left. + \frac{b_1^2(\log ^2t - \log t - 1) + b_0b_2}{b_0^4t^2} \right]
\end{eqnarray}
where $t=\log\frac{\mu^2}{\Lambda_{QCD}^2}$, $b_0=\frac{33-2n_f}{12\pi}$, $b_1=\frac{153-19n_f}{24\pi^2}$, $b_2=\frac{2857-\frac{5033}{9}n_f+\frac{325}{27}n_f^2}{128\pi^3}$, $\Lambda_{QCD}=210$ MeV, 292 MeV and 332 MeV for $n_f=5$, 4 and 3, respectively~\cite{ParticleDataGroup:2024cfk}. The minimum subtraction masses of $c$ and $b$ quarks are taken from PDG~\cite{ParticleDataGroup:2024cfk}, which are $m_c(m_c)=1.275\pm0.025$ GeV and $m_b(m_b)=4.18\pm0.03$ GeV. For the nucleons $p$, $N^{*}$ and single charmed baryon, the typical energy scale $\mu=1$ GeV is often adopted, and the energy scale $\mu=2$ GeV often works well~\cite{Wang:2020mxk} for single bottom baryon.
\subsection{The mass and pole residue of $N^*(1535)$}\label{sec4.1}
\begin{figure}
	\centering
	\includegraphics[width=8.5cm]{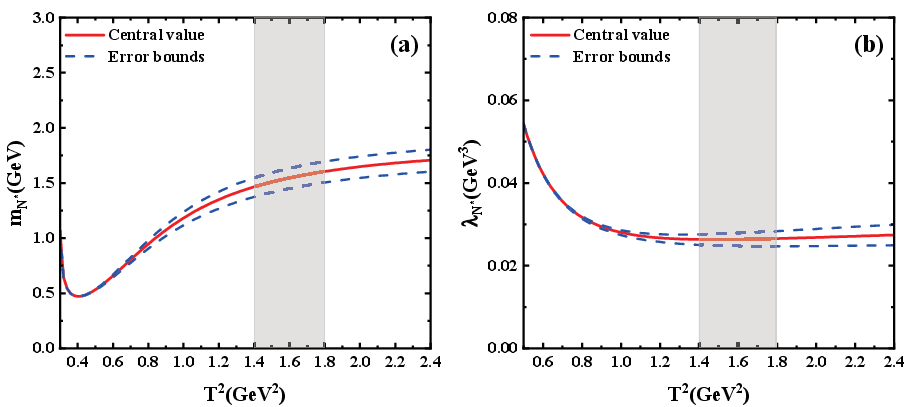}
	\caption{The mass (a) and pole residue (b) for $N^*(1535)$ with variation of the Borel parameter $T^2$, where the grey bound denote the Borel platform.}
	\label{mlambda}
\end{figure}

The mass and pole residue of $N^*(1535)$ are determined according to Eq. (\ref{eq:13}). From this equation, one can find that the results of QCD sum rules depend on the Borel parameter $T^2$ and the threshold parameter $u_0$. To obtain reliable results, an appropriate work region of Borel parameter should be selected. The hadron parameter like mass and pole residue should have a weak Borel parameter dependency in this region. This work region is commonly named as 'Borel platform'. At the same time, two conditions should be satisfied which are the pole dominance and the convergence of OPE. The pole contribution and the $n$-dimension vacuum condensate contribution can be defined as,
\begin{eqnarray}
\notag
\mathrm{PC}(T^2)=\frac{\int\limits_0^{u_0} dse^ {- u/T^2} [m_p\rho _1^{\mathrm{QCD}}(u) + \rho _2^{\mathrm{QCD}}(u)]}{\int\limits_0^{\infty} due^ {- u/T^2} [m_p\rho _1^{\mathrm{QCD}}(u) + \rho _2^{\mathrm{QCD}}(u)]}\\
D(n)=\frac{\int\limits_0^{u_0} due^ {- u/T^2} [m_p\rho _1^{\mathrm{QCD}(n)}(u) + \rho _2^{\mathrm{QCD}(n)}(u)]}{\int\limits_0^{u_0} dse^ {- u/T^2} [m_p\rho _1^{\mathrm{QCD}}(u) + \rho _2^{\mathrm{QCD}}(u)]}
\end{eqnarray}
where $\rho_i^{\mathrm{QCD}(n)}(u)$ denotes the QCD spectral density of n-dimension vacuum condensate. The pole dominance requires that the pole contribution should be larger than $50\%$, and the convergence of OPE require the contribution of high dimension condensate should be small. After repeated trial and contrast, the threshold parameter $u_0$ is determined as $4.35-5.22$ GeV$^2$, and the Borel platform is taken as $1.4-1.8$ GeV$^2$. The pole contribution in Borel platform is about $50\%$, and the contributions originate from high dimension condensates of $\left\langle {\bar qq} \right\rangle \left\langle {g_s^2GG} \right\rangle$ and $\left\langle {\bar qq} \right\rangle \left\langle {\bar q{g_s}\sigma Gq} \right\rangle $ are less than $5\%$ and $3\%$, respectively. The Borel platforms for $N^*(1535)$ are shown in Fig. \ref{mlambda}, and the final results about the mass and residue are determined to be,
\begin{eqnarray}
\notag
m_{N^*}&&=1.53^{+0.16}_{-0.16} \mathrm{GeV} \\
\lambda_{N^*}&&=0.026^{+0.002}_{-0.001} \mathrm{GeV^3}
\end{eqnarray}

\subsection{The strong coupling constants of vertices $\Lambda_cD^{(*)}N^*$ and $\Lambda_bB^{(*)}N^*$}\label{sec4.2}
\begin{figure*}
	\centering
	\includegraphics[width=16cm]{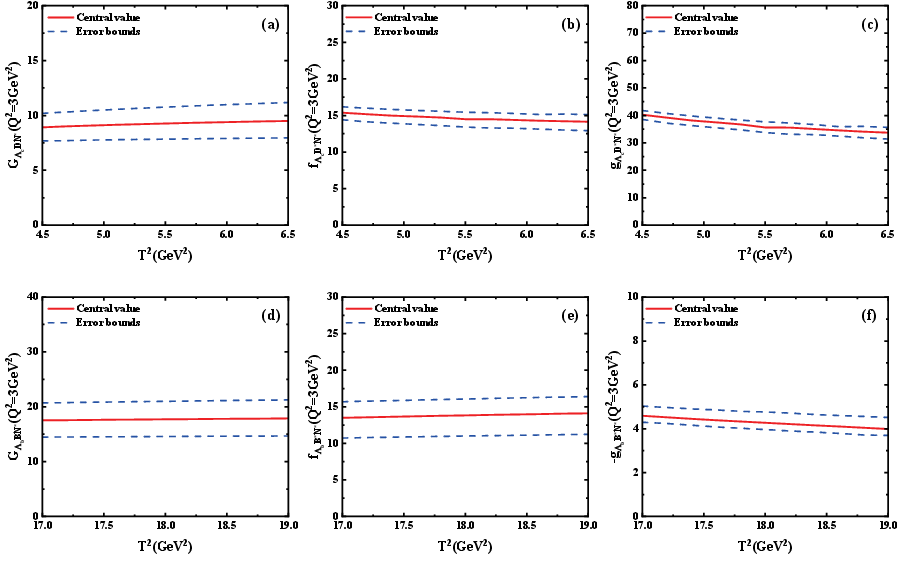}
	\caption{The Borel windows for strong coupling constants $G_{\Lambda_cDN^*}$ (a), $f_{\Lambda_cD^*N^*}$ (b), $g_{\Lambda_cD^*N^*}$ (c), $G_{\Lambda_bBN^*}$ (d), $f_{\Lambda_bB^*N^*}$ (e) and $g_{\Lambda_bB^*N^*}$ (f)  in $Q^2=3$ GeV$^2$.}
	\label{GT}
\end{figure*}

The strong coupling constants of vertices $\Lambda_cD^{(*)}N^*$ and $\Lambda_bB^{(*)}N^*$ are determined by Eq. (\ref{eq:35}). From this equation, the strong coupling constants depend on the threshold parameters $s_0$ and $u_0$, the Borel parameter $T^2$ and the square of momentum $Q^2$. The value of parameter $u_0$ has been determined to be $u_0=4.35-5.22$ GeV$^2$ in Sec.~\ref{sec4.1} according to two-point QCD sum rules. $s_{0}$ is the threshold of initial state single charmed (bottom) baryon, its value is chosen to be $s_0=7.02 (36.00)-8.12 (38.44)$ GeV$^2$ from our previous work~\cite{Wang:2020mxk}. It is similar as the procedure to analyze mass and residue, Borel platforms for strong coupling constants can also be determined according to the conditions of pole dominance and the convergence of OPE. For more details about this problem, one can consult our previous work~\cite{Lu:2023pcg}. Fixing $Q^2=3$ GeV$^2$ in Eq.~(\ref{eq:35}), the Borel platforms of these strong coupling constants are determined after repeated trial and contrast which are shown in Fig.~\ref{GT}. In these Borel platforms, the central value of pole contribution is about 40$\%$. The contributions of vacuum condensate $\left\langle g_s^2GG \right\rangle$ and $g_s^2\left\langle \bar qq \right\rangle^2$ are less than 1$\%$, which means that the convergence of OPE is also satisfied.

\begin{figure*}
	\centering
	\includegraphics[width=16cm]{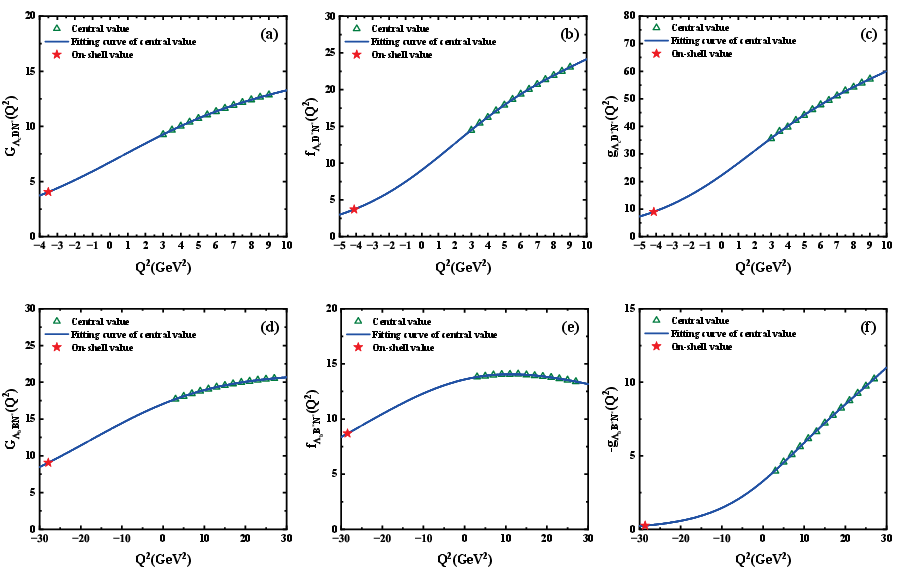}
	\caption{The fitting curves of coupling constants $G_{\Lambda_cDN^*}$ (a), $f_{\Lambda_cD^*N^*}$ (b) and $g_{\Lambda_cD^*N^*}$ (c), $G_{\Lambda_bBN^*}$ (d), $f_{\Lambda_bB^*N^*}$ (e) and $g_{\Lambda_bB^*N^*}$ (f).}
	\label{GQ}
\end{figure*}

By taking different values of $Q^2$ in space-like regions ($Q^2>0$), the momentum dependent strong coupling constants $G(Q^2)$ can be obtained. The results are explicitly shown in Fig. \ref{GQ}, where $Q^2$ are in the range of $3-9$ GeV$^2$ for vertices $\Lambda_cD^{(*)}N^*$ and $3-29$ GeV$^2$ for vertices $\Lambda_bB^{(*)}N^*$. To obtain the on-shell values of the strong coupling constants, it is necessary to extrapolate them into time-like regions ($Q^{2}<0$). This process be realized by fitting the $G(Q^{2})$ with appropriate analytical functions. Finally, the on-shell values can be obtained by setting $Q^{2}=-m_{D^{(*)}[B^{(*)}]}^2$ in analytical functions. In general, the single (multiple) pole function, exponential function, linear function or the combination of these functions are usually employed to fit the momentum dependent coupling constants~\cite{Bracco:2011pg,Wang:2013iia,Azizi:2015tya}. After repeated trial and error, we find the coupling constants for vertices $\Lambda_cD^{(*)}N^*$ and $\Lambda_bB^{(*)}N^*$ in space-like regions can be well fitted by the following analytical function,
\begin{eqnarray}\label{eq:39}
G(Q^2) = \frac{G_0}{1 +\delta_1 \frac{Q^2}{m_{\Lambda_{c[b]}}^2} +\delta_2\left(\frac{Q^2}{m_{\Lambda_{c[b]}}^2}\right)^2 }\exp\left(\delta_3\frac{Q^2}{m_{\Lambda_{c[b]}}^2}\right)
\end{eqnarray}
where $G_0$, $\delta_1$, $\delta_2$ and $\delta_3$ are the fitting parameters and their values are all listed in Table \ref{FP}. To quantify the goodness of the fitting model, we also calculate the coefficients of determination ($R$-Squared) of the fitting function for all coupling constants, which are shown in the last column of Table~\ref{FP}. It can be find that the $R$-Squared for all coupling constants are close to 1, which indicates that the fitting function is highly consistent with the numerical results. The fitting curves for the coupling constants of vertices $\Lambda_cD^{(*)}N^*$ and $\Lambda_bB^{(*)}N^*$ are shown in Fig. \ref{GQ}.
\begin{table}[htbp]
	\begin{ruledtabular}\caption{Fitting parameters and the coefficients of determination in this work.}
		\label{FP}
		\begin{tabular}{c c c c c c}
			Coupling&$G_0$&$\delta_1$&$\delta_2$&$\delta_3$&$R^2$\\ \hline
			$G_{\Lambda_{c}DN^*}$&$6.76$&$-0.09$&$0.18$&$0.56$&$0.9999$\\
			$f_{\Lambda_{c}D^*N^*}$&$9.09$&$-0.21$&$0.28$&$0.76$&$0.9999$\\
			$g_{\Lambda_{c}D^*N^*}$&$22.33$&$-0.20$&$0.28$&$0.78$&$0.9995$\\
		    $G_{\Lambda_{b}B^*N^*}$&$17.06$&$0.22$&$0.34$&$0.64$&$0.9999$\\
		    $f_{\Lambda_{b}B^*N^*}$&$13.59$&$0.16$&$0.34$&$0.36$&$0.9957$\\
		    $-g_{\Lambda_{b}B^*N^*}$&$3.27$&$-0.37$&$1.25$&$1.88$&$0.9998$\\
		\end{tabular}
	\end{ruledtabular}
\end{table}

Finally, we obtain the on-shell values of these coupling constants by setting $Q^2=-m_{D^{(*)}[B^{(*)}]}^2$ in Eq. (\ref{eq:39}). The final results are shown as,
\begin{eqnarray}\label{eq:40}
\notag
G_{\Lambda_{c}DN^*}(Q^2=-m_D^2)=4.06^{+0.96}_{-0.75}\\
\notag
f_{\Lambda_{c}D^*N^*}(Q^2=-m_{D^*}^2)=3.73^{+0.68}_{-0.16}\\
\notag
g_{\Lambda_{c}D^*N^*}(Q^2=-m_{D^*}^2)=9.22^{+3.16}_{-0.36}\\
\notag
G_{\Lambda_{b}BN^*}(Q^2=-m_B^2)=9.11^{+1.54}_{-1.61}\\
\notag
f_{\Lambda_{b}B^*N^*}(Q^2=-m_{B^*}^2)=8.55^{+2.69}_{-2.21}\\
g_{\Lambda_{b}B^*N^*}(Q^2=-m_{B^*}^2)=-0.25^{+0.16}_{-0.01}
\end{eqnarray}
In heavy quark mass limit $m_{c[b]}\to\infty$, the above strong coupling constants have the relations $G=f$ and $g=0$~\cite{Khodjamirian:2011jp}. Our predictions show that the relation $G=f$ is satisfied in both charmed and bottom section, but the relation $g=0$ is satisfied only in bottom section. This suggests that the heavy quark limit is a good approximation for the bottom quark system in the analysis of strong coupling constants. However, it may lead to larger error for the charmed quark system.

\section{Conclusions}\label{sec5}
In conclusion, we firstly analyze the mass and pole residue of the negative parity nucleon $N^*(1535)$ in the framework of two-point QCD sum rules. Taking these results as input parameters, we also analyze the momentum dependence of strong coupling constants about the vertices $\Lambda_cDN^*$, $\Lambda_cD^*N^*$, $\Lambda_bBN^*$ and $\Lambda_bB^*N^*$ within the three-point QCD sum rules. These results are obtained in deep space-like regions ($Q^2=-q^2\gg\Lambda_{QCD}^2$). To obtain on-shell values of the strong coupling constants, we fit these results into analytical functions and extrapolate them into the time-like regions ($Q^2<0$). Finally, by taking the intermediate meson on-shell, the on-shell values of these strong coupling constants are obtained. The results of strong coupling constants are important in phenomenological analysis about the strong decay behaviours of hadrons and can also help us to further understand the strong interaction between hadrons.

\section*{Acknowledgments}
The author Jie Lu thanks Prof. Zhen-Xing Zhao for valuable discussions. This work is supported by National Natural Science Foundation of China under the Grant Nos. 12175037, 12335001, 12175068, as well as supported, in part, by National Key Research and Development Program under Grant No. 2024 YFA1610504 and Natural Science Foundation of HeBei Province under the Grant No. A2024202008.

\begin{widetext}
\appendix
\section{The integral formulas for two and three Dirac delta functions.}\label{Sec:AppA}
The integral formulas of two Dirac delta functions can be expressed as,
\begin{eqnarray}
\int {{d^4}k\delta ({k^2} - m_1^2)\delta [{{(q' - k)}^2} - m_2^2]} &&= \frac{{\pi \sqrt {\lambda (r,m_1^2,m_2^2)} }}{{2r}}\\
\int {{d^4}k\delta ({k^2} - m_1^2)\delta [{{(q' - k)}^2} - m_2^2]} {k_\mu } &&= \frac{{\pi \sqrt {\lambda (r,m_1^2,m_2^2)} }}{{2r}}\frac{{r + m_1^2 - m_2^2}}{{2r}}q{'_\mu }\\
\int {{d^4}k\delta ({k^2} - m_1^2)\delta [{{(q' - k)}^2} - m_2^2]} {k_\mu }{k_\nu } &&= \frac{{\pi \sqrt {\lambda (r,m_1^2,m_2^2)} }}{{2r}}\left[ { - \frac{{\lambda (r,m_1^2,m_2^2)}}{{12r}}{g_{\mu \nu }} + \left( {m_1^2 + \frac{{\lambda (r,m_1^2,m_2^2)}}{{3r}}} \right)\frac{{q{'_\mu }q{'_\nu }}}{{q{'^2}}}} \right]
\end{eqnarray}
where, $r=q'^2$ and $\lambda(a,b,c)=a^2+b^2+c^2-2(ab+ac+bc)$ is the triangle function.

The integral formulas of three Dirac delta functions can be given as follows,
\begin{eqnarray}
\int {{d^4}k} \delta ({k^2} - m_1^2)\delta [{(k + q)^2} - m_2^2]\delta [{(p' - k)^2} - m_3^2] &&= \frac{\pi }{{2\sqrt {\lambda (s,u,{q^2})} }}\\
\int {{d^4}k} \delta ({k^2} - m_1^2)\delta [{(k + q)^2} - m_2^2]\delta [{(p' - k)^2} - m_3^2]{k_\mu } &&= \frac{\pi }{{2\sqrt {\lambda (s,u,{q^2})} }}({\alpha _1}{p_\mu } + {\beta _1}p{'_\mu })\\
\notag
\int {{d^4}k} \delta ({k^2} - m_1^2)\delta [{(k + q)^2} - m_2^2]\delta [{(p' - k)^2} - m_3^2]{k_\mu }{k_\nu } &&= \frac{\pi }{{2\sqrt {\lambda (s,u,{q^2})} }}[{\alpha _2}{g_{\mu \nu }} + {\beta _2}{p_\mu }{p_\nu } + {\gamma _2}(p{'_\mu }{p_\nu } + {p_\mu }p{'_\nu }) + {\delta _2}p{'_\mu }p{'_\nu }]\\
\end{eqnarray}
where $s=p^2$, $u=p'^2$, $q=p-p'$,
\begin{eqnarray}
\notag
{\alpha _1} &&= \frac{{( - 2m_2^2 + {q^2} + s - u)u + m_3^2({q^2} - s + u) + m_1^2( - {q^2} + s + u)}}{{\lambda (s,u,{q^2})}}\\
{\beta _1} &&= \frac{{{q^4} - 2m_1^2s - {q^2}s + m_3^2({q^2} + s - u) - 2{q^2}u - su + {u^2} + m_2^2( - {q^2} + s + u)}}{{\lambda (s,u,{q^2})}}
\end{eqnarray}
\begin{eqnarray}
\notag
{\alpha _2} &&=\frac{1}{{2\lambda (s,u,{q^2})}} \{m_1^4s + m_1^2[m_2^2({q^2} - s - u) + s( - {q^2} + s - u) - m_3^2({q^2} + s - u)] + m_2^4u+ {q^2}[m_3^4 + m_3^2({q^2} - s - u) + su]\\
\notag
&& - m_2^2[({q^2} + s - u)u + m_3^2({q^2} - s + u)]\}\\
\notag
{\beta _2} &&=\frac{1}{{\lambda {{(s,u,{q^2})}^2}}} \{m_3^4[{q^4} - 2{q^2}(s - 2u) + {{(s - u)}^2}] + [6m_2^4 + {q^4} + {q^2}(4s - 2u) + {{(s - u)}^2} - 6m_2^2({q^2} + s - u)]{u^2}\\
\notag
&&+ {m_1^4[{q^4} + {s^2} + 4su + {u^2} - 2{q^2}(s + u)] - 2m_3^2u[ - 2{q^4} + {{(s - u)}^2} + 3m_2^2({q^2} - s + u) + {q^2}(s + u)]}\\
\notag
&&- {2m_1^2m_3^2[{q^4} + {s^2} + su - 2{u^2} + {q^2}( - 2s + u)] - 2m_1^2u[{q^4} - 2{s^2} + {q^2}(s - 2u) + su + {u^2} + 3m_2^2( - {q^2} + s + u)]}\}\\
\notag
{\gamma _2} &&=-\frac{1}{\lambda (s,u,{q^2})} [{(m_1^2 - m_3^2 + u)(m_2^2 - m_3^2 - {q^2} + u) - 2m_1^2( - {q^2} + s + u)}]\\
\notag
&&-\frac{3( - {q^2} + s + u)}{\lambda {{(s,u,{q^2})}^2}} \{ m_1^4s + m_1^2[m_2^2({q^2} - s - u) + s( - {q^2} + s - u) - m_3^2({q^2} + s - u)]+ {q^2}[m_3^4 + m_3^2({q^2} - s - u) + su] \\
\notag
&& + m_2^4u- m_2^2[u({q^2} + s - u) + m_3^2({q^2} - s + u)] \}\\
\notag
{\delta _2} &&=\frac{1}{\lambda {{(s,u,{q^2})}^2}} \{q^8 - 4m_1^2{q^4}s - 2{q^6}s + 6m_1^4{s^2} + 2m_1^2{q^2}{s^2} + {q^4}{s^2} + 2m_1^2{s^3} + m_3^2[{q^4} + {q^2}(4s - 2u) + {{(s - u)}^2}] - 4{q^6}u\\
\notag
&&+ 8m_1^2{q^2}su + 2{q^4}su + 2m_1^2{s^2}u + 4{q^2}{s^2}u + 6{q^4}{u^2} - 4m_1^2s{u^2} + 2{q^2}s{u^2} + {s^2}{u^2} - 4{q^2}{u^3} - 2s{u^3} + {u^4}+ m_2^4[{q^4} + {s^2} \\
\notag
&& + 4su+u^2- 2{q^2}(s + u)] - 2m_2^2m_3^2[{q^4} - 2{s^2} + {q^2}(s - 2u) + su + {u^2}]+ 2m_3^2[{q^6} + {q^4}(s - 3u) - (s - u)(3m_1^2s + su - {u^2})\\
\notag
&& - {q^2}(3m_1^2s+ 2{s^2} + 3su - 3{u^2})] - 2m_2^2[{q^6} + 3m_1^2{s^2} + 3m_1^2su + 2{s^2}u - s{u^2} - {u^3} - {q^4}(2s + 3u)]- 2m_2^2{q^2}(- 3m_1^2s + {s^2}\\
&& + 3su + 3{u^2})\}
\end{eqnarray}
It is noted that a geometric constraint will be introduced in the integration of three Dirac delta functions, and it has the following form,
\begin{eqnarray}
- 1 \le \cos \theta  = \frac{{(u - {q^2} + m_2^2 - m_3^2)(s + u - {q^2}) - 2s(u + m_1^2 - m_3^2)}}{{\sqrt {{{(u - {q^2} + m_2^2 - m_3^2)}^2} - 4sm_1^2} \sqrt {\lambda (s,u,q^2)} }} \le 1
\end{eqnarray}

\end{widetext}

\bibliography{ref.bib}

\begin{thebibliography}{61}
\expandafter\ifx\csname natexlab\endcsname\relax\def\natexlab#1{#1}\fi
\expandafter\ifx\csname bibnamefont\endcsname\relax
  \def\bibnamefont#1{#1}\fi
\expandafter\ifx\csname bibfnamefont\endcsname\relax
  \def\bibfnamefont#1{#1}\fi
\expandafter\ifx\csname citenamefont\endcsname\relax
  \def\citenamefont#1{#1}\fi
\expandafter\ifx\csname url\endcsname\relax
  \def\url#1{\texttt{#1}}\fi
\expandafter\ifx\csname urlprefix\endcsname\relax\def\urlprefix{URL }\fi
\providecommand{\bibinfo}[2]{#2}
\providecommand{\eprint}[2][]{\url{#2}}

\bibitem[{\citenamefont{Xiao et~al.}(2019)\citenamefont{Xiao, Huang, Dong,
  Geng, and Chen}}]{Xiao:2019mvs}
\bibinfo{author}{\bibfnamefont{C.-J.} \bibnamefont{Xiao}},
  \bibinfo{author}{\bibfnamefont{Y.}~\bibnamefont{Huang}},
  \bibinfo{author}{\bibfnamefont{Y.-B.} \bibnamefont{Dong}},
  \bibinfo{author}{\bibfnamefont{L.-S.} \bibnamefont{Geng}}, \bibnamefont{and}
  \bibinfo{author}{\bibfnamefont{D.-Y.} \bibnamefont{Chen}},
  \bibinfo{journal}{Phys. Rev. D} \textbf{\bibinfo{volume}{100}},
  \bibinfo{pages}{014022} (\bibinfo{year}{2019}), \eprint{1904.00872}.

\bibitem[{\citenamefont{He and Chen}(2019)}]{He:2019rva}
\bibinfo{author}{\bibfnamefont{J.}~\bibnamefont{He}} \bibnamefont{and}
  \bibinfo{author}{\bibfnamefont{D.-Y.} \bibnamefont{Chen}},
  \bibinfo{journal}{Eur. Phys. J. C} \textbf{\bibinfo{volume}{79}},
  \bibinfo{pages}{887} (\bibinfo{year}{2019}), \eprint{1909.05681}.

\bibitem[{\citenamefont{Wu and Chen}(2024)}]{Wu:2024lud}
\bibinfo{author}{\bibfnamefont{Q.}~\bibnamefont{Wu}} \bibnamefont{and}
  \bibinfo{author}{\bibfnamefont{D.-Y.} \bibnamefont{Chen}},
  \bibinfo{journal}{Phys. Rev. D} \textbf{\bibinfo{volume}{109}},
  \bibinfo{pages}{094003} (\bibinfo{year}{2024}), \eprint{2402.14467}.

\bibitem[{\citenamefont{Han et~al.}(2021)\citenamefont{Han, Jiang, Liu, Xiao,
  and Yu}}]{Han:2021azw}
\bibinfo{author}{\bibfnamefont{J.-J.} \bibnamefont{Han}},
  \bibinfo{author}{\bibfnamefont{H.-Y.} \bibnamefont{Jiang}},
  \bibinfo{author}{\bibfnamefont{W.}~\bibnamefont{Liu}},
  \bibinfo{author}{\bibfnamefont{Z.-J.} \bibnamefont{Xiao}}, \bibnamefont{and}
  \bibinfo{author}{\bibfnamefont{F.-S.} \bibnamefont{Yu}},
  \bibinfo{journal}{Chin. Phys. C} \textbf{\bibinfo{volume}{45}},
  \bibinfo{pages}{053105} (\bibinfo{year}{2021}), \eprint{2101.12019}.

\bibitem[{\citenamefont{Jia et~al.}(2024)\citenamefont{Jia, Jiang, Wang, and
  Yu}}]{Jia:2024pyb}
\bibinfo{author}{\bibfnamefont{C.-P.} \bibnamefont{Jia}},
  \bibinfo{author}{\bibfnamefont{H.-Y.} \bibnamefont{Jiang}},
  \bibinfo{author}{\bibfnamefont{J.-P.} \bibnamefont{Wang}}, \bibnamefont{and}
  \bibinfo{author}{\bibfnamefont{F.-S.} \bibnamefont{Yu}},
  \bibinfo{journal}{JHEP} \textbf{\bibinfo{volume}{11}}, \bibinfo{pages}{072}
  (\bibinfo{year}{2024}), \eprint{2408.14959}.

\bibitem[{\citenamefont{Hu et~al.}(2025)\citenamefont{Hu, Jia, Xing, and
  Yu}}]{Hu:2024uia}
\bibinfo{author}{\bibfnamefont{X.-H.} \bibnamefont{Hu}},
  \bibinfo{author}{\bibfnamefont{C.-P.} \bibnamefont{Jia}},
  \bibinfo{author}{\bibfnamefont{Y.}~\bibnamefont{Xing}}, \bibnamefont{and}
  \bibinfo{author}{\bibfnamefont{F.-S.} \bibnamefont{Yu}},
  \bibinfo{journal}{Phys. Rev. D} \textbf{\bibinfo{volume}{111}},
  \bibinfo{pages}{076002} (\bibinfo{year}{2025}), \eprint{2403.09511}.

\bibitem[{\citenamefont{Altmeyer et~al.}(1995)\citenamefont{Altmeyer, Gockeler,
  Horsley, Laermann, Schierholz, and Zerwas}}]{Altmeyer:1995qx}
\bibinfo{author}{\bibfnamefont{R.~L.} \bibnamefont{Altmeyer}},
  \bibinfo{author}{\bibfnamefont{M.}~\bibnamefont{Gockeler}},
  \bibinfo{author}{\bibfnamefont{R.}~\bibnamefont{Horsley}},
  \bibinfo{author}{\bibfnamefont{E.}~\bibnamefont{Laermann}},
  \bibinfo{author}{\bibfnamefont{G.}~\bibnamefont{Schierholz}},
  \bibnamefont{and} \bibinfo{author}{\bibfnamefont{P.~M.}
  \bibnamefont{Zerwas}}, \bibinfo{journal}{Z. Phys. C}
  \textbf{\bibinfo{volume}{68}}, \bibinfo{pages}{443} (\bibinfo{year}{1995}),
  \eprint{hep-lat/9504003}.

\bibitem[{\citenamefont{Bracco et~al.}(1999)\citenamefont{Bracco, Navarra, and
  Nielsen}}]{Bracco:1999xe}
\bibinfo{author}{\bibfnamefont{M.~E.} \bibnamefont{Bracco}},
  \bibinfo{author}{\bibfnamefont{F.~S.} \bibnamefont{Navarra}},
  \bibnamefont{and} \bibinfo{author}{\bibfnamefont{M.}~\bibnamefont{Nielsen}},
  \bibinfo{journal}{Phys. Lett. B} \textbf{\bibinfo{volume}{454}},
  \bibinfo{pages}{346} (\bibinfo{year}{1999}), \eprint{nucl-th/9902007}.

\bibitem[{\citenamefont{Bracco and Nielsen}(2010)}]{Bracco:2010bf}
\bibinfo{author}{\bibfnamefont{M.~E.} \bibnamefont{Bracco}} \bibnamefont{and}
  \bibinfo{author}{\bibfnamefont{M.}~\bibnamefont{Nielsen}},
  \bibinfo{journal}{Phys. Rev. D} \textbf{\bibinfo{volume}{82}},
  \bibinfo{pages}{034012} (\bibinfo{year}{2010}), \eprint{1002.4990}.

\bibitem[{\citenamefont{Bracco et~al.}(2012)\citenamefont{Bracco, Chiapparini,
  Navarra, and Nielsen}}]{Bracco:2011pg}
\bibinfo{author}{\bibfnamefont{M.~E.} \bibnamefont{Bracco}},
  \bibinfo{author}{\bibfnamefont{M.}~\bibnamefont{Chiapparini}},
  \bibinfo{author}{\bibfnamefont{F.~S.} \bibnamefont{Navarra}},
  \bibnamefont{and} \bibinfo{author}{\bibfnamefont{M.}~\bibnamefont{Nielsen}},
  \bibinfo{journal}{Prog. Part. Nucl. Phys.} \textbf{\bibinfo{volume}{67}},
  \bibinfo{pages}{1019} (\bibinfo{year}{2012}), \eprint{1104.2864}.

\bibitem[{\citenamefont{Cui et~al.}(2012)\citenamefont{Cui, Liu, and
  Huang}}]{Cui:2012wk}
\bibinfo{author}{\bibfnamefont{C.-Y.} \bibnamefont{Cui}},
  \bibinfo{author}{\bibfnamefont{Y.-L.} \bibnamefont{Liu}}, \bibnamefont{and}
  \bibinfo{author}{\bibfnamefont{M.-Q.} \bibnamefont{Huang}},
  \bibinfo{journal}{Phys. Lett. B} \textbf{\bibinfo{volume}{711}},
  \bibinfo{pages}{317} (\bibinfo{year}{2012}), \eprint{1204.3979}.

\bibitem[{\citenamefont{Wang}(2014)}]{Wang:2013iia}
\bibinfo{author}{\bibfnamefont{Z.-G.} \bibnamefont{Wang}},
  \bibinfo{journal}{Phys. Rev. D} \textbf{\bibinfo{volume}{89}},
  \bibinfo{pages}{034017} (\bibinfo{year}{2014}), \eprint{1307.2422}.

\bibitem[{\citenamefont{Yu et~al.}(2015)\citenamefont{Yu, Li, and
  Wang}}]{Yu:2015xwa}
\bibinfo{author}{\bibfnamefont{G.-L.} \bibnamefont{Yu}},
  \bibinfo{author}{\bibfnamefont{Z.-Y.} \bibnamefont{Li}}, \bibnamefont{and}
  \bibinfo{author}{\bibfnamefont{Z.-G.} \bibnamefont{Wang}},
  \bibinfo{journal}{Eur. Phys. J. C} \textbf{\bibinfo{volume}{75}},
  \bibinfo{pages}{243} (\bibinfo{year}{2015}), \eprint{1502.01698}.

\bibitem[{\citenamefont{Lu et~al.}(2023{\natexlab{a}})\citenamefont{Lu, Yu, and
  Wang}}]{Lu:2023gmd}
\bibinfo{author}{\bibfnamefont{J.}~\bibnamefont{Lu}},
  \bibinfo{author}{\bibfnamefont{G.-L.} \bibnamefont{Yu}}, \bibnamefont{and}
  \bibinfo{author}{\bibfnamefont{Z.-G.} \bibnamefont{Wang}},
  \bibinfo{journal}{Eur. Phys. J. A} \textbf{\bibinfo{volume}{59}},
  \bibinfo{pages}{195} (\bibinfo{year}{2023}{\natexlab{a}}),
  \eprint{2304.13969}.

\bibitem[{\citenamefont{Lu et~al.}(2024)\citenamefont{Lu, Yu, Wang, and
  Wu}}]{Lu:2023lvu}
\bibinfo{author}{\bibfnamefont{J.}~\bibnamefont{Lu}},
  \bibinfo{author}{\bibfnamefont{G.-L.} \bibnamefont{Yu}},
  \bibinfo{author}{\bibfnamefont{Z.-G.} \bibnamefont{Wang}}, \bibnamefont{and}
  \bibinfo{author}{\bibfnamefont{B.}~\bibnamefont{Wu}}, \bibinfo{journal}{Chin.
  Phys. C} \textbf{\bibinfo{volume}{48}}, \bibinfo{pages}{013102}
  (\bibinfo{year}{2024}), \eprint{2307.05090}.

\bibitem[{\citenamefont{Colangelo and De~Fazio}(1998)}]{Colangelo:1997rp}
\bibinfo{author}{\bibfnamefont{P.}~\bibnamefont{Colangelo}} \bibnamefont{and}
  \bibinfo{author}{\bibfnamefont{F.}~\bibnamefont{De~Fazio}},
  \bibinfo{journal}{Eur. Phys. J. C} \textbf{\bibinfo{volume}{4}},
  \bibinfo{pages}{503} (\bibinfo{year}{1998}), \eprint{hep-ph/9706271}.

\bibitem[{\citenamefont{Zhu and Dai}(1998)}]{Zhu:1998vf}
\bibinfo{author}{\bibfnamefont{S.-L.} \bibnamefont{Zhu}} \bibnamefont{and}
  \bibinfo{author}{\bibfnamefont{Y.-B.} \bibnamefont{Dai}},
  \bibinfo{journal}{Phys. Rev. D} \textbf{\bibinfo{volume}{58}},
  \bibinfo{pages}{094033} (\bibinfo{year}{1998}), \eprint{hep-ph/9802225}.

\bibitem[{\citenamefont{Khodjamirian et~al.}(1999)\citenamefont{Khodjamirian,
  Ruckl, Weinzierl, and Yakovlev}}]{Khodjamirian:1999hb}
\bibinfo{author}{\bibfnamefont{A.}~\bibnamefont{Khodjamirian}},
  \bibinfo{author}{\bibfnamefont{R.}~\bibnamefont{Ruckl}},
  \bibinfo{author}{\bibfnamefont{S.}~\bibnamefont{Weinzierl}},
  \bibnamefont{and} \bibinfo{author}{\bibfnamefont{O.~I.}
  \bibnamefont{Yakovlev}}, \bibinfo{journal}{Phys. Lett. B}
  \textbf{\bibinfo{volume}{457}}, \bibinfo{pages}{245} (\bibinfo{year}{1999}),
  \eprint{hep-ph/9903421}.

\bibitem[{\citenamefont{Wang and Wan}(2006)}]{Wang:2006bs}
\bibinfo{author}{\bibfnamefont{Z.~G.} \bibnamefont{Wang}} \bibnamefont{and}
  \bibinfo{author}{\bibfnamefont{S.~L.} \bibnamefont{Wan}},
  \bibinfo{journal}{Phys. Rev. D} \textbf{\bibinfo{volume}{73}},
  \bibinfo{pages}{094020} (\bibinfo{year}{2006}), \eprint{hep-ph/0603007}.

\bibitem[{\citenamefont{Wang}(2007)}]{Wang:2007mc}
\bibinfo{author}{\bibfnamefont{Z.-G.} \bibnamefont{Wang}},
  \bibinfo{journal}{Eur. Phys. J. C} \textbf{\bibinfo{volume}{52}},
  \bibinfo{pages}{553} (\bibinfo{year}{2007}), \eprint{0705.3720}.

\bibitem[{\citenamefont{Khodjamirian et~al.}(2011)\citenamefont{Khodjamirian,
  Klein, Mannel, and Wang}}]{Khodjamirian:2011jp}
\bibinfo{author}{\bibfnamefont{A.}~\bibnamefont{Khodjamirian}},
  \bibinfo{author}{\bibfnamefont{C.}~\bibnamefont{Klein}},
  \bibinfo{author}{\bibfnamefont{T.}~\bibnamefont{Mannel}}, \bibnamefont{and}
  \bibinfo{author}{\bibfnamefont{Y.~M.} \bibnamefont{Wang}},
  \bibinfo{journal}{JHEP} \textbf{\bibinfo{volume}{09}}, \bibinfo{pages}{106}
  (\bibinfo{year}{2011}), \eprint{1108.2971}.

\bibitem[{\citenamefont{Khodjamirian et~al.}(2021)\citenamefont{Khodjamirian,
  Meli\'c, Wang, and Wei}}]{Khodjamirian:2020mlb}
\bibinfo{author}{\bibfnamefont{A.}~\bibnamefont{Khodjamirian}},
  \bibinfo{author}{\bibfnamefont{B.}~\bibnamefont{Meli\'c}},
  \bibinfo{author}{\bibfnamefont{Y.-M.} \bibnamefont{Wang}}, \bibnamefont{and}
  \bibinfo{author}{\bibfnamefont{Y.-B.} \bibnamefont{Wei}},
  \bibinfo{journal}{JHEP} \textbf{\bibinfo{volume}{03}}, \bibinfo{pages}{016}
  (\bibinfo{year}{2021}), \eprint{2011.11275}.

\bibitem[{\citenamefont{Aliev and \c{S}im\c{s}ek}(2021)}]{Aliev:2020lly}
\bibinfo{author}{\bibfnamefont{T.~M.} \bibnamefont{Aliev}} \bibnamefont{and}
  \bibinfo{author}{\bibfnamefont{K.}~\bibnamefont{\c{S}im\c{s}ek}},
  \bibinfo{journal}{Phys. Rev. D} \textbf{\bibinfo{volume}{103}},
  \bibinfo{pages}{054044} (\bibinfo{year}{2021}), \eprint{2011.07150}.

\bibitem[{\citenamefont{Aliev et~al.}(2021)\citenamefont{Aliev, Barakat, and
  \c{S}im\c{s}ek}}]{Aliev:2021hqq}
\bibinfo{author}{\bibfnamefont{T.~M.} \bibnamefont{Aliev}},
  \bibinfo{author}{\bibfnamefont{T.}~\bibnamefont{Barakat}}, \bibnamefont{and}
  \bibinfo{author}{\bibfnamefont{K.}~\bibnamefont{\c{S}im\c{s}ek}},
  \bibinfo{journal}{Eur. Phys. J. A} \textbf{\bibinfo{volume}{57}},
  \bibinfo{pages}{160} (\bibinfo{year}{2021}), \eprint{2101.10264}.

\bibitem[{\citenamefont{Aliev et~al.}(2022)\citenamefont{Aliev, Bilmis, and
  Savci}}]{Aliev:2022gxi}
\bibinfo{author}{\bibfnamefont{T.~M.} \bibnamefont{Aliev}},
  \bibinfo{author}{\bibfnamefont{S.}~\bibnamefont{Bilmis}}, \bibnamefont{and}
  \bibinfo{author}{\bibfnamefont{M.}~\bibnamefont{Savci}},
  \bibinfo{journal}{Phys. Rev. D} \textbf{\bibinfo{volume}{106}},
  \bibinfo{pages}{074022} (\bibinfo{year}{2022}), \eprint{2208.10365}.

\bibitem[{\citenamefont{Oh et~al.}(2001)\citenamefont{Oh, Song, and
  Lee}}]{Oh:2000qr}
\bibinfo{author}{\bibfnamefont{Y.-s.} \bibnamefont{Oh}},
  \bibinfo{author}{\bibfnamefont{T.}~\bibnamefont{Song}}, \bibnamefont{and}
  \bibinfo{author}{\bibfnamefont{S.~H.} \bibnamefont{Lee}},
  \bibinfo{journal}{Phys. Rev. C} \textbf{\bibinfo{volume}{63}},
  \bibinfo{pages}{034901} (\bibinfo{year}{2001}), \eprint{nucl-th/0010064}.

\bibitem[{\citenamefont{Li et~al.}(2002)\citenamefont{Li, Huang, Sun, and
  Dai}}]{Li:2002pp}
\bibinfo{author}{\bibfnamefont{Z.-H.} \bibnamefont{Li}},
  \bibinfo{author}{\bibfnamefont{T.}~\bibnamefont{Huang}},
  \bibinfo{author}{\bibfnamefont{J.-Z.} \bibnamefont{Sun}}, \bibnamefont{and}
  \bibinfo{author}{\bibfnamefont{Z.-H.} \bibnamefont{Dai}},
  \bibinfo{journal}{Phys. Rev. D} \textbf{\bibinfo{volume}{65}},
  \bibinfo{pages}{076005} (\bibinfo{year}{2002}), \eprint{hep-ph/0208168}.

\bibitem[{\citenamefont{Deandrea et~al.}(2003)\citenamefont{Deandrea, Nardulli,
  and Polosa}}]{Deandrea:2003pv}
\bibinfo{author}{\bibfnamefont{A.}~\bibnamefont{Deandrea}},
  \bibinfo{author}{\bibfnamefont{G.}~\bibnamefont{Nardulli}}, \bibnamefont{and}
  \bibinfo{author}{\bibfnamefont{A.~D.} \bibnamefont{Polosa}},
  \bibinfo{journal}{Phys. Rev. D} \textbf{\bibinfo{volume}{68}},
  \bibinfo{pages}{034002} (\bibinfo{year}{2003}), \eprint{hep-ph/0302273}.

\bibitem[{\citenamefont{Shifman
  et~al.}(1979{\natexlab{a}})\citenamefont{Shifman, Vainshtein, and
  Zakharov}}]{Shifman:1978bx}
\bibinfo{author}{\bibfnamefont{M.~A.} \bibnamefont{Shifman}},
  \bibinfo{author}{\bibfnamefont{A.~I.} \bibnamefont{Vainshtein}},
  \bibnamefont{and} \bibinfo{author}{\bibfnamefont{V.~I.}
  \bibnamefont{Zakharov}}, \bibinfo{journal}{Nucl. Phys. B}
  \textbf{\bibinfo{volume}{147}}, \bibinfo{pages}{385}
  (\bibinfo{year}{1979}{\natexlab{a}}).

\bibitem[{\citenamefont{Shifman
  et~al.}(1979{\natexlab{b}})\citenamefont{Shifman, Vainshtein, and
  Zakharov}}]{Shifman:1978by}
\bibinfo{author}{\bibfnamefont{M.~A.} \bibnamefont{Shifman}},
  \bibinfo{author}{\bibfnamefont{A.~I.} \bibnamefont{Vainshtein}},
  \bibnamefont{and} \bibinfo{author}{\bibfnamefont{V.~I.}
  \bibnamefont{Zakharov}}, \bibinfo{journal}{Nucl. Phys. B}
  \textbf{\bibinfo{volume}{147}}, \bibinfo{pages}{448}
  (\bibinfo{year}{1979}{\natexlab{b}}).

\bibitem[{\citenamefont{Novikov et~al.}(1978)\citenamefont{Novikov, Okun,
  Shifman, Vainshtein, Voloshin, and Zakharov}}]{Novikov:1977dq}
\bibinfo{author}{\bibfnamefont{V.~A.} \bibnamefont{Novikov}},
  \bibinfo{author}{\bibfnamefont{L.~B.} \bibnamefont{Okun}},
  \bibinfo{author}{\bibfnamefont{M.~A.} \bibnamefont{Shifman}},
  \bibinfo{author}{\bibfnamefont{A.~I.} \bibnamefont{Vainshtein}},
  \bibinfo{author}{\bibfnamefont{M.~B.} \bibnamefont{Voloshin}},
  \bibnamefont{and} \bibinfo{author}{\bibfnamefont{V.~I.}
  \bibnamefont{Zakharov}}, \bibinfo{journal}{Phys. Rept.}
  \textbf{\bibinfo{volume}{41}}, \bibinfo{pages}{1} (\bibinfo{year}{1978}).

\bibitem[{\citenamefont{Ioffe}(1981)}]{Ioffe:1981kw}
\bibinfo{author}{\bibfnamefont{B.~L.} \bibnamefont{Ioffe}},
  \bibinfo{journal}{Nucl. Phys. B} \textbf{\bibinfo{volume}{188}},
  \bibinfo{pages}{317} (\bibinfo{year}{1981}), \bibinfo{note}{[Erratum:
  Nucl.Phys.B 191, 591--592 (1981)]}.

\bibitem[{\citenamefont{Narison}(1987)}]{Narison:1987qc}
\bibinfo{author}{\bibfnamefont{S.}~\bibnamefont{Narison}},
  \bibinfo{journal}{Phys. Lett. B} \textbf{\bibinfo{volume}{198}},
  \bibinfo{pages}{104} (\bibinfo{year}{1987}).

\bibitem[{\citenamefont{Matheus et~al.}(2009)\citenamefont{Matheus, Navarra,
  Nielsen, and Zanetti}}]{Matheus:2009vq}
\bibinfo{author}{\bibfnamefont{R.~D.} \bibnamefont{Matheus}},
  \bibinfo{author}{\bibfnamefont{F.~S.} \bibnamefont{Navarra}},
  \bibinfo{author}{\bibfnamefont{M.}~\bibnamefont{Nielsen}}, \bibnamefont{and}
  \bibinfo{author}{\bibfnamefont{C.~M.} \bibnamefont{Zanetti}},
  \bibinfo{journal}{Phys. Rev. D} \textbf{\bibinfo{volume}{80}},
  \bibinfo{pages}{056002} (\bibinfo{year}{2009}), \eprint{0907.2683}.

\bibitem[{\citenamefont{Mo et~al.}(2014)\citenamefont{Mo, Cui, Liu, and
  Huang}}]{Mo:2014nua}
\bibinfo{author}{\bibfnamefont{Z.}~\bibnamefont{Mo}},
  \bibinfo{author}{\bibfnamefont{C.-Y.} \bibnamefont{Cui}},
  \bibinfo{author}{\bibfnamefont{Y.-L.} \bibnamefont{Liu}}, \bibnamefont{and}
  \bibinfo{author}{\bibfnamefont{M.-Q.} \bibnamefont{Huang}},
  \bibinfo{journal}{Commun. Theor. Phys.} \textbf{\bibinfo{volume}{61}},
  \bibinfo{pages}{501} (\bibinfo{year}{2014}), \eprint{1403.6906}.

\bibitem[{\citenamefont{Wang}(2015)}]{Wang:2015mxa}
\bibinfo{author}{\bibfnamefont{Z.-G.} \bibnamefont{Wang}},
  \bibinfo{journal}{Eur. Phys. J. C} \textbf{\bibinfo{volume}{75}},
  \bibinfo{pages}{427} (\bibinfo{year}{2015}), \eprint{1506.01993}.

\bibitem[{\citenamefont{Wang and Wang}(2021)}]{Wang:2020mxk}
\bibinfo{author}{\bibfnamefont{Z.-G.} \bibnamefont{Wang}} \bibnamefont{and}
  \bibinfo{author}{\bibfnamefont{H.-J.} \bibnamefont{Wang}},
  \bibinfo{journal}{Chin. Phys. C} \textbf{\bibinfo{volume}{45}},
  \bibinfo{pages}{013109} (\bibinfo{year}{2021}), \eprint{2006.16776}.

\bibitem[{\citenamefont{Yang and Chen}(2024)}]{Yang:2023fsc}
\bibinfo{author}{\bibfnamefont{H.-M.} \bibnamefont{Yang}} \bibnamefont{and}
  \bibinfo{author}{\bibfnamefont{H.-X.} \bibnamefont{Chen}},
  \bibinfo{journal}{Phys. Rev. D} \textbf{\bibinfo{volume}{109}},
  \bibinfo{pages}{036032} (\bibinfo{year}{2024}), \eprint{2311.01991}.

\bibitem[{\citenamefont{Zeng et~al.}(2025)\citenamefont{Zeng, Wu, Hu, Fu, and
  Zhong}}]{Zeng:2025gft}
\bibinfo{author}{\bibfnamefont{L.}~\bibnamefont{Zeng}},
  \bibinfo{author}{\bibfnamefont{X.-G.} \bibnamefont{Wu}},
  \bibinfo{author}{\bibfnamefont{D.-D.} \bibnamefont{Hu}},
  \bibinfo{author}{\bibfnamefont{H.-B.} \bibnamefont{Fu}}, \bibnamefont{and}
  \bibinfo{author}{\bibfnamefont{T.}~\bibnamefont{Zhong}},
  \bibinfo{journal}{Phys. Rev. D} \textbf{\bibinfo{volume}{111}},
  \bibinfo{pages}{056030} (\bibinfo{year}{2025}), \eprint{2501.06737}.

\bibitem[{\citenamefont{Wang et~al.}(2008)\citenamefont{Wang, Zou, Wei, Li, and
  Lu}}]{Wang:2007ys}
\bibinfo{author}{\bibfnamefont{Y.-M.} \bibnamefont{Wang}},
  \bibinfo{author}{\bibfnamefont{H.}~\bibnamefont{Zou}},
  \bibinfo{author}{\bibfnamefont{Z.-T.} \bibnamefont{Wei}},
  \bibinfo{author}{\bibfnamefont{X.-Q.} \bibnamefont{Li}}, \bibnamefont{and}
  \bibinfo{author}{\bibfnamefont{C.-D.} \bibnamefont{Lu}},
  \bibinfo{journal}{Eur. Phys. J. C} \textbf{\bibinfo{volume}{54}},
  \bibinfo{pages}{107} (\bibinfo{year}{2008}), \eprint{0707.1138}.

\bibitem[{\citenamefont{Azizi et~al.}(2015{\natexlab{a}})\citenamefont{Azizi,
  Sarac, and Sundu}}]{Azizi:2015tya}
\bibinfo{author}{\bibfnamefont{K.}~\bibnamefont{Azizi}},
  \bibinfo{author}{\bibfnamefont{Y.}~\bibnamefont{Sarac}}, \bibnamefont{and}
  \bibinfo{author}{\bibfnamefont{H.}~\bibnamefont{Sundu}},
  \bibinfo{journal}{Nucl. Phys. A} \textbf{\bibinfo{volume}{943}},
  \bibinfo{pages}{159} (\bibinfo{year}{2015}{\natexlab{a}}),
  \eprint{1501.05084}.

\bibitem[{\citenamefont{Azizi et~al.}(2015{\natexlab{b}})\citenamefont{Azizi,
  Sarac, and Sundu}}]{Azizi:2015bxa}
\bibinfo{author}{\bibfnamefont{K.}~\bibnamefont{Azizi}},
  \bibinfo{author}{\bibfnamefont{Y.}~\bibnamefont{Sarac}}, \bibnamefont{and}
  \bibinfo{author}{\bibfnamefont{H.}~\bibnamefont{Sundu}},
  \bibinfo{journal}{Phys. Rev. D} \textbf{\bibinfo{volume}{92}},
  \bibinfo{pages}{014022} (\bibinfo{year}{2015}{\natexlab{b}}),
  \eprint{1506.00809}.

\bibitem[{\citenamefont{Yu et~al.}(2017)\citenamefont{Yu, Wang, and
  Li}}]{Yu:2016pyo}
\bibinfo{author}{\bibfnamefont{G.~L.} \bibnamefont{Yu}},
  \bibinfo{author}{\bibfnamefont{Z.~G.} \bibnamefont{Wang}}, \bibnamefont{and}
  \bibinfo{author}{\bibfnamefont{Z.~Y.} \bibnamefont{Li}},
  \bibinfo{journal}{Chin. Phys. C} \textbf{\bibinfo{volume}{41}},
  \bibinfo{pages}{083104} (\bibinfo{year}{2017}), \eprint{1608.03460}.

\bibitem[{\citenamefont{Yu et~al.}(2019)\citenamefont{Yu, Guan, and
  Wang}}]{Yu:2018hnv}
\bibinfo{author}{\bibfnamefont{G.-L.} \bibnamefont{Yu}},
  \bibinfo{author}{\bibfnamefont{R.-H.} \bibnamefont{Guan}}, \bibnamefont{and}
  \bibinfo{author}{\bibfnamefont{Z.-G.} \bibnamefont{Wang}},
  \bibinfo{journal}{Int. J. Mod. Phys. A} \textbf{\bibinfo{volume}{33}},
  \bibinfo{pages}{1850217} (\bibinfo{year}{2019}), \eprint{1810.05970}.

\bibitem[{\citenamefont{Shi et~al.}(2020)\citenamefont{Shi, Wang, and
  Zhao}}]{Shi:2019hbf}
\bibinfo{author}{\bibfnamefont{Y.-J.} \bibnamefont{Shi}},
  \bibinfo{author}{\bibfnamefont{W.}~\bibnamefont{Wang}}, \bibnamefont{and}
  \bibinfo{author}{\bibfnamefont{Z.-X.} \bibnamefont{Zhao}},
  \bibinfo{journal}{Eur. Phys. J. C} \textbf{\bibinfo{volume}{80}},
  \bibinfo{pages}{568} (\bibinfo{year}{2020}), \eprint{1902.01092}.

\bibitem[{\citenamefont{Zhao et~al.}(2020)\citenamefont{Zhao, Li, Shen, Shi,
  and Yang}}]{Zhao:2020mod}
\bibinfo{author}{\bibfnamefont{Z.-X.} \bibnamefont{Zhao}},
  \bibinfo{author}{\bibfnamefont{R.-H.} \bibnamefont{Li}},
  \bibinfo{author}{\bibfnamefont{Y.-L.} \bibnamefont{Shen}},
  \bibinfo{author}{\bibfnamefont{Y.-J.} \bibnamefont{Shi}}, \bibnamefont{and}
  \bibinfo{author}{\bibfnamefont{Y.-S.} \bibnamefont{Yang}},
  \bibinfo{journal}{Eur. Phys. J. C} \textbf{\bibinfo{volume}{80}},
  \bibinfo{pages}{1181} (\bibinfo{year}{2020}), \eprint{2010.07150}.

\bibitem[{\citenamefont{Zhao et~al.}(2023)\citenamefont{Zhao, Sun, Zhang, Xing,
  and Yang}}]{Zhao:2021sje}
\bibinfo{author}{\bibfnamefont{Z.-X.} \bibnamefont{Zhao}},
  \bibinfo{author}{\bibfnamefont{X.-Y.} \bibnamefont{Sun}},
  \bibinfo{author}{\bibfnamefont{F.-W.} \bibnamefont{Zhang}},
  \bibinfo{author}{\bibfnamefont{Y.-P.} \bibnamefont{Xing}}, \bibnamefont{and}
  \bibinfo{author}{\bibfnamefont{Y.-T.} \bibnamefont{Yang}},
  \bibinfo{journal}{Phys. Rev. D} \textbf{\bibinfo{volume}{108}},
  \bibinfo{pages}{116008} (\bibinfo{year}{2023}), \eprint{2103.09436}.

\bibitem[{\citenamefont{Zhang and Qiao}(2023)}]{Zhang:2023nxl}
\bibinfo{author}{\bibfnamefont{S.-Q.} \bibnamefont{Zhang}} \bibnamefont{and}
  \bibinfo{author}{\bibfnamefont{C.-F.} \bibnamefont{Qiao}},
  \bibinfo{journal}{Phys. Rev. D} \textbf{\bibinfo{volume}{108}},
  \bibinfo{pages}{074017} (\bibinfo{year}{2023}), \eprint{2307.05019}.

\bibitem[{\citenamefont{Lu et~al.}(2023{\natexlab{b}})\citenamefont{Lu, Yu,
  Wang, and Wu}}]{Lu:2023pcg}
\bibinfo{author}{\bibfnamefont{J.}~\bibnamefont{Lu}},
  \bibinfo{author}{\bibfnamefont{G.-L.} \bibnamefont{Yu}},
  \bibinfo{author}{\bibfnamefont{Z.-G.} \bibnamefont{Wang}}, \bibnamefont{and}
  \bibinfo{author}{\bibfnamefont{B.}~\bibnamefont{Wu}}, \bibinfo{journal}{Eur.
  Phys. J. C} \textbf{\bibinfo{volume}{83}}, \bibinfo{pages}{907}
  (\bibinfo{year}{2023}{\natexlab{b}}), \eprint{2308.06705}.

\bibitem[{\citenamefont{Wu et~al.}(2024)\citenamefont{Wu, Yu, Lu, and
  Wang}}]{Wu:2024gcq}
\bibinfo{author}{\bibfnamefont{B.}~\bibnamefont{Wu}},
  \bibinfo{author}{\bibfnamefont{G.-L.} \bibnamefont{Yu}},
  \bibinfo{author}{\bibfnamefont{J.}~\bibnamefont{Lu}}, \bibnamefont{and}
  \bibinfo{author}{\bibfnamefont{Z.-G.} \bibnamefont{Wang}},
  \bibinfo{journal}{Phys. Lett. B} \textbf{\bibinfo{volume}{859}},
  \bibinfo{pages}{139118} (\bibinfo{year}{2024}), \eprint{2406.08181}.

\bibitem[{\citenamefont{Zhang et~al.}(2024)\citenamefont{Zhang, Zhang, and
  Qiao}}]{Zhang:2024ick}
\bibinfo{author}{\bibfnamefont{S.-Q.} \bibnamefont{Zhang}},
  \bibinfo{author}{\bibfnamefont{X.-H.} \bibnamefont{Zhang}}, \bibnamefont{and}
  \bibinfo{author}{\bibfnamefont{C.-F.} \bibnamefont{Qiao}},
  \bibinfo{journal}{JHEP} \textbf{\bibinfo{volume}{06}}, \bibinfo{pages}{122}
  (\bibinfo{year}{2024}), \eprint{2402.15088}.

\bibitem[{\citenamefont{Lu et~al.}(2025)\citenamefont{Lu, Chen, Yu, Wang, and
  Wu}}]{Lu:2025bvi}
\bibinfo{author}{\bibfnamefont{J.}~\bibnamefont{Lu}},
  \bibinfo{author}{\bibfnamefont{D.-Y.} \bibnamefont{Chen}},
  \bibinfo{author}{\bibfnamefont{G.-L.} \bibnamefont{Yu}},
  \bibinfo{author}{\bibfnamefont{Z.-G.} \bibnamefont{Wang}}, \bibnamefont{and}
  \bibinfo{author}{\bibfnamefont{B.}~\bibnamefont{Wu}}, \bibinfo{journal}{Phys.
  Rev. D} \textbf{\bibinfo{volume}{111}}, \bibinfo{pages}{114037}
  (\bibinfo{year}{2025}), \eprint{2501.15534}.

\bibitem[{\citenamefont{Pascual and Tarrach}(1984)}]{Pascual:1984zb}
\bibinfo{author}{\bibfnamefont{P.}~\bibnamefont{Pascual}} \bibnamefont{and}
  \bibinfo{author}{\bibfnamefont{R.}~\bibnamefont{Tarrach}},
  \emph{\bibinfo{title}{{QCD: RENORMALIZATION FOR THE PRACTITIONER}}}, vol.
  \bibinfo{volume}{194} (\bibinfo{year}{1984}).

\bibitem[{\citenamefont{Reinders et~al.}(1985)\citenamefont{Reinders,
  Rubinstein, and Yazaki}}]{Reinders:1984sr}
\bibinfo{author}{\bibfnamefont{L.~J.} \bibnamefont{Reinders}},
  \bibinfo{author}{\bibfnamefont{H.}~\bibnamefont{Rubinstein}},
  \bibnamefont{and} \bibinfo{author}{\bibfnamefont{S.}~\bibnamefont{Yazaki}},
  \bibinfo{journal}{Phys. Rept.} \textbf{\bibinfo{volume}{127}},
  \bibinfo{pages}{1} (\bibinfo{year}{1985}).

\bibitem[{\citenamefont{Navas et~al.}(2024)}]{ParticleDataGroup:2024cfk}
\bibinfo{author}{\bibfnamefont{S.}~\bibnamefont{Navas}} \bibnamefont{et~al.}
  (\bibinfo{collaboration}{Particle Data Group}), \bibinfo{journal}{Phys. Rev.
  D} \textbf{\bibinfo{volume}{110}}, \bibinfo{pages}{030001}
  (\bibinfo{year}{2024}).

\bibitem[{\citenamefont{Yu et~al.}(2021)\citenamefont{Yu, Wang, Wang, and
  Wang}}]{Yu:2021ggd}
\bibinfo{author}{\bibfnamefont{G.-L.} \bibnamefont{Yu}},
  \bibinfo{author}{\bibfnamefont{Z.-G.} \bibnamefont{Wang}},
  \bibinfo{author}{\bibfnamefont{X.-W.} \bibnamefont{Wang}}, \bibnamefont{and}
  \bibinfo{author}{\bibfnamefont{H.-J.} \bibnamefont{Wang}},
  \bibinfo{journal}{Int. J. Mod. Phys. A} \textbf{\bibinfo{volume}{36}},
  \bibinfo{pages}{2150197} (\bibinfo{year}{2021}), \eprint{2102.11078}.

\bibitem[{\citenamefont{Cutkosky}(1960)}]{Cutkosky:1960sp}
\bibinfo{author}{\bibfnamefont{R.~E.} \bibnamefont{Cutkosky}},
  \bibinfo{journal}{J. Math. Phys.} \textbf{\bibinfo{volume}{1}},
  \bibinfo{pages}{429} (\bibinfo{year}{1960}).

\bibitem[{\citenamefont{Leljak et~al.}(2021)\citenamefont{Leljak, Meli\'c, and
  van Dyk}}]{Leljak:2021vte}
\bibinfo{author}{\bibfnamefont{D.}~\bibnamefont{Leljak}},
  \bibinfo{author}{\bibfnamefont{B.}~\bibnamefont{Meli\'c}}, \bibnamefont{and}
  \bibinfo{author}{\bibfnamefont{D.}~\bibnamefont{van Dyk}},
  \bibinfo{journal}{JHEP} \textbf{\bibinfo{volume}{07}}, \bibinfo{pages}{036}
  (\bibinfo{year}{2021}), \eprint{2102.07233}.

\bibitem[{\citenamefont{Narison}(2010)}]{Narison:2010cg}
\bibinfo{author}{\bibfnamefont{S.}~\bibnamefont{Narison}},
  \bibinfo{journal}{Phys. Lett. B} \textbf{\bibinfo{volume}{693}},
  \bibinfo{pages}{559} (\bibinfo{year}{2010}), \bibinfo{note}{[Erratum:
  Phys.Lett.B 705, 544--544 (2011)]}, \eprint{1004.5333}.

\bibitem[{\citenamefont{Narison}(2012{\natexlab{a}})}]{Narison:2011xe}
\bibinfo{author}{\bibfnamefont{S.}~\bibnamefont{Narison}},
  \bibinfo{journal}{Phys. Lett. B} \textbf{\bibinfo{volume}{706}},
  \bibinfo{pages}{412} (\bibinfo{year}{2012}{\natexlab{a}}),
  \eprint{1105.2922}.

\bibitem[{\citenamefont{Narison}(2012{\natexlab{b}})}]{Narison:2011rn}
\bibinfo{author}{\bibfnamefont{S.}~\bibnamefont{Narison}},
  \bibinfo{journal}{Phys. Lett. B} \textbf{\bibinfo{volume}{707}},
  \bibinfo{pages}{259} (\bibinfo{year}{2012}{\natexlab{b}}),
  \eprint{1105.5070}.

\end{thebibliography}

\end{document}